\documentclass[12pt]{article}

\pdfoutput=1

\usepackage{color}

\usepackage{amsmath,amssymb,latexsym}

\usepackage{cite}

\usepackage[T1]{fontenc}
\usepackage{times}

\usepackage{epstopdf}

\usepackage{graphics,epsfig}

\usepackage{caption}
\usepackage{fdsymbol}

\usepackage{indentfirst}

\let\a=\alpha \let\b=\beta \let\g=\gamma \let\d=\delta \let\e=\epsilon
\let\z=\zeta  
 \let\k=\kappa
\let\l=\lambda \let\m=\mu \let\n=\nu \let\x=\xi \let\p=\pi 
\let\s=\sigma 

\let\w=\omega        \let\Th=\Theta 
\let\X=\Xi  \let\S=\Sigma  \let\Y=\Psi
 \let\W=\Omega
\let\la=\label  
  
 \def\bd{\begin{document}} \def\ed{\end{document}}
\def\ds{\documentstyle} \let\fr=\frac \let\bl=\bigl \let\br=\bigr
\let\Br=\Bigr \let\Bl=\Bigl
\let\bm=\bibitem
\let\na=\nabla
\def\tU{{\widetilde U}}
\let\pa=\partial \let\ov=\overline
\def\ie{{\it i.e.\ }}
\newcommand{\be}{\begin{equation}}
\newcommand{\ee}{\end{equation}}
\def\ba{\begin{array}}
\def\ea{\end{array}}

\def\bei{\begin{itemize}}
\def\eei{\end{itemize}}

\def\ben{\begin{enumerate}}
\def\een{\end{enumerate}}

\def\ft#1#2{{\textstyle{{\scriptstyle #1}\over {\scriptstyle #2}}}}
\def\fft#1#2{{#1 \over #2}}
\def\F#1#2{{ F_{#1}^{(#2)} }}
\def\cF#1#2{{ {\cal F}_{#1}^{(#2)} }}

\def\R{{\bf R}}
\def\sst#1{{\scriptscriptstyle #1}}
\def\oneone{\rlap 1\mkern4mu{\rm l}}
\def\e7{E_{7(+7)}}
\def\td{\tilde}
\def\wtd{\widetilde}
\def\im{{\rm i}}
\def\bog{Bogomol'nyi\ }
\newcommand{\ho}[1]{$\, ^{#1}$}
\newcommand{\hoch}[1]{$\, ^{#1}$}
\newcommand{\bea}{\begin{eqnarray}}
\newcommand{\eea}{\end{eqnarray}}
\newcommand{\ra}{\rightarrow}
\newcommand{\lra}{\longrightarrow}
\newcommand{\Lra}{\Leftrightarrow}
\newcommand{\ap}{\alpha^\prime}
\newcommand{\bp}{\tilde \beta^\prime}
\newcommand{\cB}{{\cal B}}
\newcommand{\cO}{{\cal O}}
\newcommand{\vecx}{\vec{x}}
\newcommand{\vecy}{\vec{y}}
\newcommand{\vecp}{\vec{p}}
\newcommand{\vecq}{\vec{q}}
\newcommand{\tr}{{\rm tr} }
\newcommand{\Tr}{{\rm Tr} }
\newcommand{\NP}{Nucl. Phys. }

\newcommand{\cL}{{\cal L}}
\newcommand{\cA}{{\cal A}}
\newcommand{\cT}{{\cal T}}
\newcommand{\cD}{{\cal D}}
\newcommand{\cH}{{\cal H}}

\def\th{\theta}

\def\sst#1{{\scriptscriptstyle #1}}
\def\0{{\sst{(0)}}}
\def\1{{\sst{(1)}}}
\def\2{{\sst{(2)}}}
\def\3{{\sst{(3)}}}
\def\4{{\sst{(4)}}}
\def\5{{\sst{(5)}}}
\def\6{{\sst{(6)}}}
\def\7{{\sst{(7)}}}
\def\8{{\sst{(8)}}}
\def\9{{\sst{(9)}}}
\def\p{{\sst{(p)}}}
\def\q{{\sst{(q)}}}

\def\ssa{{\sst{(\alpha)}}}
\def\ssb{{\sst{(\beta)}}}
\def\ssg{{\sst{(\gamma)}}}
\def\j{{\sst{(j)}}}

\def\ve{\varepsilon}
\def\vf{\varphi}
\def\F{\Phi}
\def\wg{\wedge}

\def\thb{\bar{\theta}}
\def\Thb{\bar{\Theta}}
\def\barp{\bar{p}}
\def\barq{\bar{q}}
\def\barc{\bar{c}}
\def\bard{\bar{d}}
\def\e{\epsilon}

\def \bi{\bibitem}
\def \la {\label}

\def \l {\lambda}
\def\foot{\footnote}
\def \tl  {{\tilde \l}}
\def \sql {{\sqrt \l}}
\def \adss {$AdS_5 \times S^5$\ }
\newcommand{\rf}[1]{(\ref{#1})}
\def \ov {\over}

\def\Th{\Theta}
\def\vth{\vartheta}
\def\btheta{{\bar\theta}}
\def\ttheta{{{\tilde\theta}}}
\def\bttheta{{{\bar\ttheta}}}
\def\vth{\vartheta}

\def\ra{\rightarrow}
\def\N{{\cal N}}
\def\uM{\underline{M}}
\def\uA{\underline{A}}
\def\uN{\underline{N}}
\def\uP{\underline{P}}
\def\ua{\underline{a}}
\def\ub{\underline{b}}
\def\uc{\underline{c}}
\def\ud{\underline{d}}
\def\ue{\underline{e}}
\def\uf{\underline{f}}
\def\ui{\underline{i}}
\def\uj{\underline{j}}
\def\uk{\underline{k}}
\def\ual{\underline{\alpha}}
\def\ube{\underline{\beta}}
\def\um{\underline{m}}
\def\un{\underline{n}}
\def\up{\underline{p}}
\def\uq{\underline{q}}
\def\ur{\underline{r}}
\def\us{\underline{s}}

\def\umu{\underline{\mu}}
\def\unu{\underline{\nu}}
\def\ula{\underline{\l}}
\def\uka{\underline{\k}}
\def\usi{\underline{\s}}
\def\urh{\underline{\r}}
\def\cc{\circ}
\def\eqv{\equiv}

\def\ni{\noindent}

\def\Ep{E^{{}^{(+)}}}
\def\Em{E^{{}^{(-)}}}

\def\Mp{M^{{}^{(+)}}}
\def\Mm{M^{{}^{(-)}}}

\def \ha{{1\ov 2}}

\def\r{\rho}

\def\Y{{\rm Y}}
\def\X{{\rm X}}
\def\tY{\tilde{\rm Y}}
\def\tX{\tilde{\rm X}}
\def\dY{\dot{\rm Y}}
\def\dX{\dot{\rm X}}

\def \J {\mathcal{J}}
\def \del {\partial}

\def\dF{\dot{F}}
\def\dG{\dot{G}}
\def\df{\dot{f}}
\def\dx{\dot{x}}
\def \E {{\cal E}}
\def \S {{\cal S}}
\def \J {{\cal J}}

\def\ms{\mathcal{S}}
\def\mj{\mathcal{J}}
\def\soj{\fr{\ms}{\mj}}
\def \R {{\bf R}}
\def \om {\omega}
\def \bE {\bar E}
\def \x {{\cal X}}

\def \bi{\bibitem}
\def \la {\label}

\def \l {\lambda}
\def\foot{\footnote}
\def \tl  {{\tilde \l}}
\def \sql {{\sqrt \l}}
\def \adss {$AdS_5 \times S^5$\ }
\def \ov {\over}

\def \varpi {{\rm w}}

\def\thb{\bar{\theta}}
\def\Thb{\bar{\Theta}}
\def\mb{\bar{\m}}
\def\ab{\bar{\a}}
\def\zb{\bar{z}}
\def\psib{\bar{\psi}}
\def\barp{\bar{p}}
\def\barq{\bar{q}}
\def\barc{\bar{c}}
\def\bard{\bar{d}}
\def\e{\epsilon}
\def\wb{\bar{w}}
\def\lb{\bar{\l}}
\def\Jb{\bar{J}}
\def\Nb{\bar{N}}
\def\Zb{\bar{Z}}
\def\pab{\bar{\pa}}

\def\At{\tilde{A}}
\def\Bt{\tilde{B}}
\def\Ct{\tilde{C}}
\def\Dt{\tilde{D}}
\def\Et{\tilde{E}}
\def\Ft{\tilde{F}}
\def\Gt{\tilde{G}}
\def\Ht{\tilde{H}}
\def\Mt{\tilde{M}}
\def\Rt{\tilde{R}}
\def\at{\tilde{a}}
\def\bt{\tilde{b}}
\def\ct{\tilde{c}}
\def\dt{\tilde{d}}
\def\et{\tilde{e}}
\def\ft{\tilde{f}}
\def\gt{\tilde{g}}
\def\mt{\tilde{\mu}}
\def\nt{\tilde{\nu}}

\def\asth{\hat{*}}
\def\phh{\hat{\phi}}

\def\bA{{\bf A}}

\def\ola{\overleftarrow}
\def\ora{\overrightarrow}
\def\alt{\tilde{\a}}

\def\eh{\hat{e}}
\def\eph{\hat{\e}}
\def\ph{\hat{p}}
\def\alh{\hat{\a}}
\def\beh{\hat{\b}}
\def\gah{\hat{\g}}
\def\Fh{\hat{F}}
\def\muh{\hat{\m}}
\def\nuh{\hat{\n}}
\def\thh{\hat{\th}}
\def\dh{\hat{d}}
\def\ih{\hat{i}}
\def\jh{\hat{j}}
\def\kh{\hat{k}}
\def\deh{\hat{\d}}
\def\wh{\hat{w}}
\def\lah{\hat{\l}}
\def\Ah{\hat{A}}
\def\Ch{\hat{C}}
\def\Omh{\hat{\Omega}}

\def\xh{\hat{x}}

\def\ps{\rlap{\, /}\;\,p }
\def\ks{\rlap{\, /}\;\,k }

\def\gym{g_{YM}}

\def\adot{\dot{a}}
\def\bdot{\dot{b}}
\def\bpa{\bar{\pa}}

\def\pr{\prime}
\def\ssk{\medskip}
\def\bsk{\bigskip}

\def\clb{\color{blue}}
\def\clr{\color{red}}
\def\clv{\color{violet}}

\def\t{\tau}
\def\cM{\mathcal{M}}
\def\S{\Sigma}

\def\N{\nabla}

\def\cR{\mathcal{R}}
\def\cL{\mathcal{L}}

\def\hb{\hbar}
\def\an{\hat{a}}
\def\ac{\hat{a}^\dag}
\def\hp{\hat{p}}

\def\Ec{{\cal E}}

\DeclareFontFamily{U}{FdSymbolA}{}
\DeclareFontShape{U}{FdSymbolA}{m}{n}{
    <-> s * [1] FdSymbolA-Book
}{}
\DeclareFontShape{U}{FdSymbolA}{m}{b}{
    <-> s * [1] FdSymbolA-Medium
}{}
\DeclareSymbolFont{fdsymbols}{U}{FdSymbolA}{m}{n}
\SetSymbolFont{fdsymbols}{bold}{U}{FdSymbolA}{m}{b}

\DeclareMathSymbol{\medtriangleright}{\mathbin}{fdsymbols}{86}
\DeclareMathSymbol{\medtriangleup}{\mathbin}{fdsymbols}{87}
\DeclareMathSymbol{\medtriangleleft}{\mathbin}{fdsymbols}{88}
\DeclareMathSymbol{\nabla}{\mathbin}{fdsymbols}{89}

\begin{document}

\title{
{\Large{\bf 
Kerr Black Holes within the Membrane Paradigm 
}}
}
{
\author{A.M. Arslanaliev$\,^{\vardiamondsuit,\spadesuit}$\footnote{arslanaliev.kh@gmail.com}
\,\, \,and  \,A.J. Nurmagambetov$\,^{\spadesuit,\vardiamondsuit,\varheartsuit}$\footnote{ajn@kipt.kharkov.ua}
\\ \\
$\,^{\vardiamondsuit}${ \it {\normalsize Department of Physics \& Technology, Karazin Kharkiv National University,}}\\
{ \it {\normalsize 4 Svobody Sq., Kharkiv 61022, Ukraine} }
\\
$\,^{\spadesuit}${ \it {\normalsize Akhiezer Institute for Theoretical Physics of NSC KIPT}}\\
{ \it {\normalsize 1 Akademicheskaya St., Kharkiv 61108, Ukraine} }
\\
$\,^{\varheartsuit}${ \it {\normalsize Usikov Institute of Radiophysics and Electronics}}\\
{ \it {\normalsize 12 Ak. Proskury, Kharkiv 61085, Ukraine} }
}

\date{}

\maketitle

\abstract{
We consider the membrane viewpoint a l\`a Parikh-Wilczek on the Kerr solution for a rotating black hole. Computing the stress-energy tensor of a close-to-the-horizon stretched membrane and comparing it to the stress-tensor of a viscous fluid, we recover transport coefficients in terms of the Kerr geometry. Viscosities of the dual fluid remain constant, while the rest of the transport coefficients become complex functions of radial and angle coordinates. We study the qualitative behavior of the pressure, expansion, and energy/momentum densities for two specific black holes: the slowly rotating black hole, with the angular momentum of one percent of the black hole mass squared, and the extremal Kerr black hole. For the Kerr solution in the Boyer-Lindquist coordinates, these transport coefficients generally have poles at different values of the radial coordinate in the range between the horizon and the Schwarzschild radius of the black hole, in dependence on the fixed angle direction. We briefly discuss our findings in the context of a relation between the Membrane Paradigm and the AdS/CFT correspondence, the KSS bound violation, the coordinate choice, and a non-stationary extension of the Kerr solution. 
}

\newpage

\section{Introduction}

This work is aimed at the application of the Membrane Paradigm within the Parikh-Wilczek approach to rotating black holes. Developing pioneering researches on Electrodynamics and Gravitodynamics near black hole horizons \cite{Damour:1978cg,Damour:1982}, the Membrane Paradigm \cite{Thorne:1986iy,Price:1986yy,Suen:1988kq,Parikh:1997ma, Chatterjee:2010gp} appeals to the effective dynamics of a viscous fluid, simulated characteristics of a stretched membrane located at the horizon. Later on, the Duality between the hydrodynamic limit of a strongly coupled Gauge Theory and weakly coupled Gravity with the matter was used for computing transport coefficients of the Gauge Theory on the Gravity side \cite{Kovtun:2004de}. Surprisingly, some of the results of the ``old'' Membrane Paradigm and of the ``new'' Gauge/Gravity Duality matched in the early development of the latter (see, e.g., \cite{Iqbal:2008by,Kovtun:2008kx,Ritz:2010zza}). Notwithstanding, the Membrane Paradigm does not refer anyway to conformal symmetry, which plays a significant role as in the original AdS/CFT correspondence, as well as in its ``light'' version, known as the Gauge/Gravity Duality.

\ssk
To date, the strong differentiation between the results of computation of shear viscosity on both~-- the Membrane Paradigm and the Gauge/Gravity Duality -- sides was not determined. Considering this rationale as a guideline, before doing non-trivial computations in the Gauge/Gravity Duality on the Kerr spacetime, we applied technically more simple Membrane Paradigm to compute this and other transport coefficients of the dual fluid. In the course of our studies, we recovered old results for the shear and bulk viscosities, which are constants in the Kerr geometry, and got new upshots on the functional dependence on the radial and angle coordinates for other transport coefficients. The shear viscosity value coincides with that of the Schwarzschild geometry, which, in particular, holds up the ``shear viscosity/entropy density'' bound $\eta/s=\hbar/4\pi k_B$, known as the KSS bound \cite{Kovtun:2004de}. The dual fluid pressure, expansion, energy, and momentum densities, being functions of radial and polar angle coordinates, show different properties in dependence on the considered case. For a slowly rotating black hole, these quantities may be regular in the range from the location of circular photon orbits to spatial infinity, as it happens in the case of the black hole whose angular momentum value is one percent of the mass squared. For the extremal Kerr black hole solution in the Boyer-Lindquist coordinates, the mentioned quantities display complex behavior. In particular, all of them have poles, which in specific cases cause the flipping of the sign in the vicinity of the singular point. 

\ssk
To set up the stage for such conclusions, in Section 2, we consider the 1+1+2 split of the Kerr metric similar to that of \cite{Parikh:1997ma,Chatterjee:2010gp}, early developed for non-rotating black holes. Note that the standard 1+3 separation of coordinates has been employed to rotating black holes as in the framework of the Membrane Paradigm and in the fluid/gravity correspondence as well (see, e.g., \cite{Bhattacharyya:2007vs,Moghaddas:2012qx,Fischler:2015kro,Cropp:2016ajh,Moghaddas:2017itv,Lemos:2017lhe,Lysov:2017cmc,Rahman:2021kwb}). The explicit 1+1+2 decomposition of metric for rotating black holes a l\`a Parikh-Wilczek has not been considered in the literature, to the extent we know.

\ssk
Computing the external curvature of a 3D stretched membrane hypersurface in the vicinity of the black hole horizon, we form the membrane stress-energy tensor. Next, comparing this tensor to that of a viscous fluid, we derive the correspondence between gravitational geometric quantities and transport coefficients of the dual fluid. We recover in this way the energy density, pressure, expansion, and momentum density of the dual medium in terms of the geometric characteristics of the Kerr solution. Specifically, we find that the effects of rotation do not impact on the shear and bulk viscosity, the values of which for the Kerr black hole coincide, within the Membrane Paradigm, with that of a static neutral black hole of the same mass. What concerns the other transport coefficients is that they, as we have mentioned, become functions of radial and angular coordinates that we explicitly obtain. 

\ssk
In Section 3, we analyze the functional dependence of the corresponding quantities on the radial coordinate by fixing the angle direction in the equatorial/pole planes of either extremal (fast rotating) or non-extremal slowly rotating Kerr black hole. Here, we establish the previously announced properties of the pressure, expansion, and energy/momentum densities of the dual fluid: they are smooth functions in the range of the radial coordinate from the circular photon orbits in the equatorial plane (as the reference point) to spatial infinity for the slowly rotating black hole of the mass $M$ and of the angular momentum $J=M^2/100$. For other relations between $J$ and $M^2$ and for other polar angle planes, these quantities generally have poles. This feature becomes manifest in the extremal Kerr black hole case, where all the mentioned characteristics of the dual fluid are divergent at the Schwarzschild radius in the equatorial plane. 

\ssk
The summary of the results, their discussion, and our conclusions are collected in the last section. We also include, for the sake of completeness, two technical Appendices. There, we reproduce the relevant to the Schwarzschild black hole computations of characteristics of the dual to the stretched membrane viscous fluid, as well as the dimensional analysis of quantities we deal with.

\ssk
{\it Notation and conventions.} Here we use the ``mostly plus'' signature of the metric and $c=1$ units. The black hole's effective mass is scaled with the gravitational constant: $M=mG$, where $m$ is the usual mass of the object. Indices of small letters of the Latin alphabet -- $a,b,c,\dots$ -- run from $0$ to $3$; indices of capital letters -- $A,B,C,\dots$ -- are two-dimensional ones (of the angular subspace of the Kerr spacetime). BH is used as the shortening of a ``black hole''.

\section{Viscous fluid description of the Kerr stretched horizon} 

We begin with the standard form of the Kerr metric \cite{Kerr:1963ud} in the Boyer-Lindquist coordinates \cite{Boyer:1966qh}:
\be
ds^2=-F^2_t dt^2-2\w F_t \,dt d\vf+F^2_r dr^2+\r^2(r,\th) d\th^2+F^2_\vf d\vf^2,
\la{Kerrds2}
\ee
where we have introduced the following functions of $(r,\th)$:
\be
F_t=\left(1-\fr{2Mr}{\r^2(r,\th)}\right)^{1/2},\qquad \r^2(r,\th)=r^2+a^2 \cos^2\th,
\la{Ftrhodef}
\ee
\be
\w=-\fr{2Mr}{\r^2(r,\th)}\,a\sin^2\th \cdot F^{-1}_t,\qquad F_r=\fr{\r(r,\th)}{\Delta^{1/2}},\qquad \Delta=r^2-2Mr+a^2,
\la{fFrtridef}
\ee
\be
F_\vf=\left((r^2+a^2)\sin^2\th+\fr{2Mr}{\r^2(r,\th)}\,a^2 \sin^4\th \right)^{1/2}.
\la{Fvfdef}
\ee

\ssk
To employ the 1+1+2 Parikh-Wilczek \cite{Parikh:1997ma} decomposition of the metric, we present the line element \rf{Kerrds2} in the following equivalent form:
\be
ds^2=-\left(F_t dt+\w d\vf \right)^2+F_r^2 dr^2+\r^2(r,\th) d\th^2+\left(F^2_\vf+\w^2 \right) d\vf^2,
\la{Kerreqvds2}
\ee
so that $ds^2=\left(-u_au_b+n_a n_b+\g_{ab} \right)dx^a dx^b$ is realized by 1-forms
\be
u_a dx^a=F_t dt+\w d\vf,\qquad n_a dx^a=F_r dr,
\la{uanaKerr}
\ee
and by the metric of a 2D subspace of angular coordinates
\be
\g_{ab}\, dx^a dx^b=\r^2(r,\th) d\th^2+\left(F^2_\vf+\w^2\right) d\vf^2,\quad \g_{ab}=g_{ab}-n_an_b+u_au_b.
\la{gabKerr}
\ee
In dual basis, vectors $u^a$ and $n^a$ have the components
\be
u^a=-F^{-1}_t \pa_t, \qquad n^a=F^{-1}_r \pa_r,
\la{unupKerr}
\ee
and the norm of $u^a$ and $n^a$ is $-1$ and $+1$, respectively.

\ssk
The $n_a$ field acceleration $n^b \nabla_b n_a$ becomes non-trivial in the case. (Indirectly, it means the presence of a non-trivial H\'aji\v{c}ek field \cite{Hajicek:1974oua}, see below.) However, one can still use the standard expression of \cite{Parikh:1997ma} for a stretched membrane stress-tensor  
\be
8\pi G t_{ab}=Kh_{ab}-K_{ab},\qquad h_{ab}=g_{ab}-n_an_b,
\la{STtensor}
\ee
even in this case (see Appendix A for details). The 3D extrinsic curvature tensor of the orthogonal to $n^a$ hypersurface is
\be
K_{ab}={h_a}^c \nabla_c n_b=\fr1{2F_r}\left(
\begin{array}{cccc}
-\pa_r F^2_t &0&0&-\pa_r(\w F_t)\\
0&0&0&0\\
0&0&\pa_r \r^2 &0\\
-\pa_r(\w F_t) &0&0&\pa_r F^2_\vf
\end{array}
\right).
\la{KabKerr}
\ee
The trace of this tensor comes to
\be
K\equiv \Tr\,K_{ab}=\fr1{2F_r}\pa_r \ln\left[\r^2(r,\th)F^2_t \left(\w^2+F^2_\vf\right)\right],
\la{KKerr}
\ee
or, in terms of the standard for the Kerr solution variables,
\be
K=\fr{1}{\Delta^{1/2} \r^3(r,\th)}\left(\r^2(r,\th) (r-M)+r\Delta \right).
\la{KKerr2}
\ee

\ssk
Now, following \cite{Parikh:1997ma}, we have to 
decouple the time-like direction and to present the 3D extrinsic curvature $K_{ab}$ in terms of 2D extrinsic curvature $k_{AB}$ of the normal to $u^a$ hypersurface. Clearly, the metric on this hypersurface is a $4 \times 4$ matrix $\g_{ab}$, the only non-trivial part of which is the $2\times 2$ block $\g_{AB}$ determined by eq. \rf{gabKerr}. Details on the relation between 3D and 2D extrinsic curvatures can be found in Appendix A. Here we just recall that $K_{ab}=\a^{-1}\left(k_{ab}-g_{\cH} u_a u_b \right)+\W_a u_b+\W_b u_a$, where $\a$ is the renormalization factor making divergent quantities finite on the event horizon; $\W_a$ is the H\'aji\v{c}ek field and $g_{\cH}$ is the surface gravity. (Actually, we will not use this relation for $K_{ab}$ in what follows, so the explicit form of the H\'aji\v{c}ek field does not matter.) 
 
\ssk
In our case, we choose the renormalization factor $\a=F^{-1}_r$ with $F_r(r,\th)$ from eq. \rf{fFrtridef}, which makes it possible to form the null-vector $l^a=F^{-1}_r n^a$. (The norm of this vector is $l^a l_a=F^{-2}_r=\Delta/\r^2(r,\th)$, and it is equal to zero on the BH horizon $r_h$, defined by $\Delta(r_h)=0$.) Then, we can compute the 2D extrinsic curvature tensor as the Lie derivative of the 2D metric $\g_{AB}$ along the null-vector $l^a$. We get
\be
k_{AB}=\fr1{2F^2_r}\left(
\begin{array}{cc}
\pa_r \r^2(r,\th) & 0\\
0 & \pa_r \left(\w^2+F^2_\vf\right)
\end{array}
\right).
\la{kABKerr}
\ee
Separating the traceless part of $k_{AB}$, we write the 2D extrinsic curvature as the combination of the 2D shear tensor $\s_{AB}$ and the expansion $\Th$: $k_{AB}=\s_{AB}+1/2\, \Th\g_{AB}$, $\Tr\, \s_{AB}=0$. 

\ssk
On account of the diagonal structure of the 2D metric (cf. eq. \rf{gabKerr}), we have 3 equations to identify $\s_{AB}$ and $\Th$:

\be
\left.
\begin{aligned}
k_{\th\th}=\s_{\th\th}+\fr12 \Th \g_{\th\th},\\
k_{\vf\vf}=\s_{\vf\vf}+\fr12 \Th \g_{\vf\vf}
\end{aligned}\right\}
\ee
and $\g^{\th\th}\s_{\th\th}+\g^{\vf\vf}\s_{\vf\vf}=0$. Therefore,
the expansion is defined by
\be
\Th=\g^{AB} k_{AB}=\fr1{2 F^2_r}\pa_r \ln \left(\r^2(r,\th) \left(\w^2+F^2_\vf \right)\right),
\la{ThKerr}
\ee
or, in terms of the standard for the Kerr spacetime variables,
\[
\Th=\fr{\Delta}{\r^4(r,\th)}\,r+\fr{4a^2 M^2 r \sin^2\th}{\r^4(r,\th)(\r^2(r,\th)-2Mr)}\left(1-\fr{Mr}{\r^2(r,\th)}\right)\left(1-\fr{2r^2}{\r^2(r,\th)}\right)
\]
\be
+\fr1{\r^2(r,\th)}\left(1-\fr{2Mr}{\r^2(r,\th)} \right)\left(r+\fr{a^2 M \sin^2\th}{\r^2(r,\th)}\left(1-\fr{2r^2}{\r^2(r,\th)}\right)\right).
\la{ThKerr2}
\ee

\ssk
Next, taking eqs. \rf{kABKerr}-\rf{ThKerr} into account, we obtain the following non-trivial components of the shear tensor $\s_{AB}$:
\be
\s_{\th\th}=\fr{\r^2(r,\th)}{4F^2_r}\, \pa_r \ln \left(\fr{\r^2(r,\th)}{\w^2+F^2_\vf} \right)= -\fr{a^2 \sin^2\th (M-r)}{2(\r^2(r,\th)-2Mr)},
\la{sththKerr}
\ee
\be
\s_{\vf\vf}=\fr{\w^2+F^2_\vf}{4F^2_r}\, \pa_r \ln \left(\fr{\w^2+F^2_\vf}{\r^2(r,\th)} \right)=\fr{a^2 \sin^4 \th \Delta (M-r)}{2(\r^2(r,\th)-2Mr)^2}. 
\la{svfvfKerr}
\ee

\ssk
Now, we are ready to move on to solving the main task: to derive transport coefficients of the dual fluid in terms of the Kerr geometry variables. Let us turn to the stretched membrane stress-tensor $8\pi G t_{ab}=Kh_{ab}-K_{ab}$, which is (cf. eqs. \rf{KabKerr}, \rf{KKerr})
\be
8\pi G t_{ab}=\left(
\begin{array}{cccc}
-K F^2_t+\fr1{2F_r}\pa_r F^2_t &0&0& -K\w F_t+\fr1{2F_r}\pa_r(\w F_t)\\
0&0&0&0\\
0&0&K\r^2(r,\th)-\fr1{2F_r}\pa_r \r^2(r,\th) &0\\
-K\w F_t+\fr1{2F_r}\pa_r(\w F_t) &0&0&K F^2_\vf-\fr1{2F_r} \pa_r F^2_\vf
\end{array}
\right).
\la{tabKerr}
\ee
To derive the transport coefficients, we have to compare the stress-tensor \rf{tabKerr} to the stress-tensor of a viscous fluid. The latter has the following general form (see, e.g., \cite{Misner:1964}, \cite{RezzollaBook}):
\be
t_{ab}=\fr1{\a} \r u_a u_b+\fr1{\a}\g_{aA}\g_{bB} \left(p \g^{AB}-2\eta \s^{AB}-\zeta \Th \g^{AB} \right)+\pi^A\left(\g_{aA}u_b+\g_{bA}u_a \right).
\la{tabfluiddef0}
\ee
Recall that indices $a,b$ are indices of the 4-dimensional spacetime; indices $A,B$ are two-dimensional indices of a surface parameterized by angles $\{\th,\vf\}$. $\r$ is the energy density, $p$ is the fluid pressure, and $\s_{AB}$ and $\Th$ are the shear tensor and the expansion of the null geodesics near the event horizon. Also, $\eta$ is the shear viscosity, $\zeta$ is the bulk viscosity, and $\pi^A$ is the momentum density. As we have noticed, the regularization factor is chosen to be $\a=1/F_r$.

\ssk
Comparing two stress-tensors from eqs. \rf{tabKerr}, \rf{tabfluiddef0}, we derive the corresponding dynamical characteristics of the fluid in terms of the Kerr geometry. For $tt$ components of both stress tensors, we have
\[
t_{tt}=\fr1{\a} \r u_t u_t \qquad \leadsto \qquad \Big|\a=F^{-1}_r,\,u_t=F_t\Big| \qquad \leadsto
\]
\be
\r=\fr1{8\pi G}\left(-\fr{K}{F_r}+\fr1{2F^2_r}\,\pa_r \ln F^2_t \right).
\la{rhoKerr}
\ee
Equivalently,
\be
\r=\fr1{8\pi G \r^4(r,\th)}\left[\fr{M \Delta \left(r^2-a^2\cos^2 \th\right)}{\r^2(r,\th)-2Mr }-\left(\r^2(r,\th) (r-M)+r \Delta \right) \right].
\la{rhoKerr2}
\ee

\ssk
Next, we have to recover the momentum density $\pi^A$ in the dual viscous fluid from the non-diagonal elements of the stretched membrane stress-energy tensor. Clearly, comparing \rf{tabKerr} and \rf{tabfluiddef0}, we get $\pi^\th=0$, and
\be
\pi^\vf=\fr1{8\pi G}\fr1{F_t\left(\w^2+F_\vf^2\right)}\left(-K\w F_t+\fr1{2F_r}\pa_r \left(\w F_t \right) \right) -\frac{F_r\w}{\w^2+F_{\vf}^2}\,\rho .
\la{pivfKerr}
\ee
With the energy density \rf{rhoKerr}, it becomes
\be
\pi^\vf= \frac{1}{16\pi G}\frac{\w}{F_r(\w^2+F_\vf^2)}\,\pa_r \ln\frac{\w}{F_t} .
\la{pivfKerr2}
\ee
Therefore, the 2D momentum density vector is $\pi^A=(\pi^\th,\pi^\vf)=(0,\pi^\vf)$ with
\be
\pi^\vf = \frac{1}{8\pi G}\frac{aM(r^2-a^2\cos^2\theta)}{\rho^4 \Delta^{1/2}(\rho^2(r,\theta)-2Mr)^{1/2}}.
\la{pivfKerr3}
\ee 

\ssk
The next components of \rf{tabKerr} we focus on are:
\be
t_{\th\th}=\fr1{8\pi G} \left(K \r^2(r,\th)-\fr1{2F_r} \pa_r \r^2(r,\th)\right),
\la{tththKerr}
\ee
\be
t_{\vf\vf}=\fr1{8\pi G} \left(K F^2_\vf-\fr1{2F_r} \pa_r F^2_\vf  \right).
\la{tvfvfKerr}
\ee
On the other hand, see eq. \rf{tabfluiddef0},
\be
t_{\th\th}=\fr1{\a}\g_{\th\th} \left(p-\zeta \Th \right)-\fr1{\a} 2\eta \s_{\th\th},
\la{tththfluid}
\ee
\be
t_{\vf\vf} = \fr1{\a}\g_{\vf\vf}\left(p-\zeta \Th \right)-\fr1{\a} 2\eta \s_{\vf\vf}+\fr1{\a}\r u_\vf u_\vf  + 2\pi^{\vf}\g_{\vf\vf}u_\vf .
\la{tvfvffluid}
\ee
Extracting the $(p-\zeta\Th)$ combination from eqs. \rf{tththfluid}, \rf{tvfvffluid} and equating the results we get
\be
2\eta = \frac{1}{F_r(\s_{\th\th}\g_{\vf\vf}-\s_{\vf\vf}\g_{\th\th})}\left( t_{\vf\vf}\g_{\th\th}-t_{\th\th}\g_{\vf\vf} -\r F_r u^2_\vf\g_{\th\th} -2\pi^\vf\g_{\vf\vf}\g_{\th\th}u_\vf \right). 
\la{2etaKerr}
\ee
Or, by use of eqs. \rf{uanaKerr}, \rf{gabKerr}, \rf{sththKerr}, \rf{svfvfKerr}, \rf{rhoKerr}, \rf{pivfKerr2}, \rf{tththKerr}, \rf{tvfvfKerr}, 
\be
\eta=\fr1{16\pi G}\left[1+\fr{\pa_r \w^2-2K \w^2 F_r-2\w^2\pa_r\ln\frac{\w}{F_t}}{(\w^2+F^2_\vf)\,\pa_r \ln \left(\r^2(r,\th)/(\w^2+F^2_\vf)\right)}\right]-\fr{\r \w^2 F^2_r}{(\w^2+F^2_\vf)\,\pa_r \ln \left(\r^2(r,\th)/(\w^2+F^2_\vf)\right)}.
\la{etaKerr2}
\ee
With eq. \rf{KKerr}, one simplifies the shear viscosity to
\be
\eta = \frac{1}{16\pi G}.
\la{etaKerr3}
\ee
Hence, we recover the standard classical result \cite{Damour:1982,Thorne:1986iy,Price:1986yy}: the shear viscosity of the dual to a Kerr BH effective fluid is constant, and it is the same as for the Schwarzschild BH. (See Appendix A.) 

\ssk
Next, multiplying both sides of \rf{tththfluid} and \rf{tvfvffluid} with $\g^{\th\th}$ and $\g^{\vf\vf}$, respectively, and using the traceless of the shear tensor $\s_{AB}$, we get
\be
p-\zeta\Theta = -\frac12 \r\g^{\vf\vf}u_\vf^2 - \frac1{F_r}\pi^\vf u_\vf+\frac1{2F_r}\left(\g^{\th\th}t_{\th\th} + \g^{\vf\vf}t_{\vf\vf}\right).
\la{pdefKerr}
\ee
As a matter of fact, the last term on the r.h.s. of \rf{pdefKerr} turns into
\be
\fr1{2 F_r} \left(\g^{\th\th} t_{\th\th}+\g^{\vf\vf} t_{\vf\vf} \right)=-\fr1{16\pi G} \Th+\fr{1}{16\pi G F_r} \left(2K-\fr{K\w^2}{\w^2+F^2_\vf}+\fr{\pa_r \w^2}{2F_r(\w^2+F^2_\vf)}\right)
\la{I2Kerr}
\ee
with the expansion $\Th$ of \rf{ThKerr}, so that
\be
p = \frac{1}{8\pi G}\frac{K}{F_r} + \left(\z - \fr1{16\pi G}\right)\Th .
\la{pKerr0}
\ee
To fix the value of the bulk viscosity, we will require the equality of the fluid pressure \rf{pKerr0}, in the limit $a\ra 0$, to the pressure of the dual to the corresponding Schwarzschild BH fluid (cf. eqs. \rf{pzeta}, \rf{ThsiggHSS} in Appendix A). Taking this limit on account of eqs. \rf{KKerr}, \rf{ThKerr}, one should get
\be
\lim_{a\ra 0} p=\fr1{8\pi G}\left(\fr12 \pa_r f+\fr{2f}{r} \right)+\left(\zeta-\fr1{16\pi G} \right)\fr{2f}{r}=\fr{1}{16\pi G}\, \pa_r f,\quad f=1-2M/r .
\la{pKerr1}
\ee
It becomes possible for
\be
\zeta=-\fr1{16\pi G}.
\la{zetaKerr}
\ee
Therefore, the bulk viscosity of the dual to a Kerr BH fluid is constant, and it coincides with the corresponding quantity in the Schwarzschild geometry \cite{Damour:1982,Thorne:1986iy,Price:1986yy}. (See also Appendix A.)

\ssk
With this choice of the bulk viscosity, the pressure, eq. \rf{pKerr0}, is reduced to
\be
p = \frac1{8\pi G}\left(\frac{K}{F_r}-\Th\right) = \frac{1}{8\pi G}\frac{1}{F_r^2}\pa_r\ln F_t,
\la{pKerr2}
\ee
or
\be
p=\fr1{8\pi G} \fr{M\Delta\left(r^2-a^2\cos^2\th \right)}{\r^4(r,\th)\left(\r^2(r,\th)-2M r \right)}.
\la{pKerr3}
\ee
This is the last transport coefficient in a viscous fluid stress-tensor \rf{tabfluiddef0}.

\ssk
To summarize the intermediate results of the work, we have established the correspondence between the Kerr geometry and transport coefficients of the dual to the event horizon stretched membrane viscous fluid. We have recovered the standard for the Membrane Paradigm results for the shear and bulk viscosities, and have explicitly obtained the functional dependence of the other transport coefficients on the radial and polar angle coordinates. In the next section, we make an analysis of these functions for two cases: of fast and slowly rotating black holes.

\section{Comments on pressure, expansion, and energy/momentum densities of the dual fluid}

Now let us review the behavior of the transport coefficients which are functions of coordinates. We begin with the fluid pressure determined by eq. \rf{pKerr3}. This expression is convenient to analyze in terms of the Schwarzschild radius $r_s=2M$:
\be
p(r,\th)=\fr1{16\pi G} \fr{r_s \left(r^2-r_s r+a^2 \right)\left(r^2-a^2 \cos^2 \th \right)}{\left(r^2-r_s r+a^2 \cos^2\th \right)\left(r^2+a^2 \cos^2 \th \right)^2}.
\la{pKerr4}
\ee
Next, we recall that the horizons of a Kerr BH are located at
\be
r^+_{\cH}=\fr12 \left(r_s+\sqrt{r^2_s-4a^2}\right),\qquad r^-_{\cH}=\fr12 \left(r_s-\sqrt{r^2_s-4a^2}\right).
\la{Kerrhor}
\ee
Apparently, the parameter $a$, related to the BH angular momentum, falls into the range
\be
0 \le \fr{a}{M} \le 1.
\la{arest}
\ee
Introducing $C\equiv a/M$, the BH horizons are determined by
\be
r_{\cH}^\pm=\fr{r_s}{2}\left(1\pm \sqrt{1-C^2} \right).
\la{Kerrhor1}
\ee

\ssk
From now on, we have to specify the type of black hole from the point of view of the intensity of its rotation. We start from the high end of the $C$ range, corresponding to the so-called extremal Kerr black hole, when the two horizons coincide: $C=1$ and $r^{extr.}_{\cH}=r_s/2=M$. Also, we have two different circular photon orbits in the equatorial plane of the Kerr black hole, which are located at \cite{Bardeen:1972fi,Teo:2020sey}
\be
r_1=r_s\left[1+\cos\left(\fr23 \arccos \left(-\fr{|a|}{M}\right)\right)\right],
\la{photon1}
\ee
\be
r_2=r_s\left[1+\cos\left(\fr23 \arccos \left(\fr{|a|}{M}\right)\right)\right].
\la{photon2}
\ee
These radii coincide at $C=|a|/M=0$ and obey $r_1 \le r_2$. According to the direction of their rotation, these two photon rings are divided into prograde (in the same rotation direction as the black hole; for $r_1$) and retrograde (in the opposite direction; for $r_2$). Fixing $C=1$ (which corresponds to the case of a fast rotating BH), we find $r_1=M$ and $r_2=4M$. We attract the reader's attention to the fact that eqs. \rf{photon1} and \rf{photon2} describe stable photon orbits in the equatorial plane; in any other plane, the location of photon orbits is not determined by \rf{photon1}, \rf{photon2}, and these orbits are generally unstable (see \cite{Teo:2020sey}). We will use $r_{1,2}$ values as the reference points in our discussion of quantities in non-equatorial planes, in particular, in the (north) pole plane.  

\ssk
Fixing the polar angle $\th=\pi/2$ and the Schwarzschild radius $r_s=1$, we get the following dependence of the rescaled by $16\pi G$ pressure on the radial coordinate for the extremal Kerr BH, Fig.\ref{fig:image1}. In this case, the radial location of the prograde photon orbits coincides with the BH horizon, $r_1=r_s/2=r^{extr.}_{\cH}$. We note that near the double extremal horizon location, equal to the Schwarzschild radius $r_s$, the pressure diverges and undergoes the sign flip. On both sides of the singular point $16\pi G p(r,\pi/2)$ is a decaying function. For the retrograde photon orbits, located far from the point of singularity, the pressure is a smooth decaying function: from $9/32 \times 1/16\pi G$ at $r_2=4r^{extr.}_{\cH}$ to $1/r^2$ at the spatial infinity.

\begin{figure}[ht]
\centering
\captionsetup{width=.8\linewidth}
\includegraphics[width=0.5\linewidth]{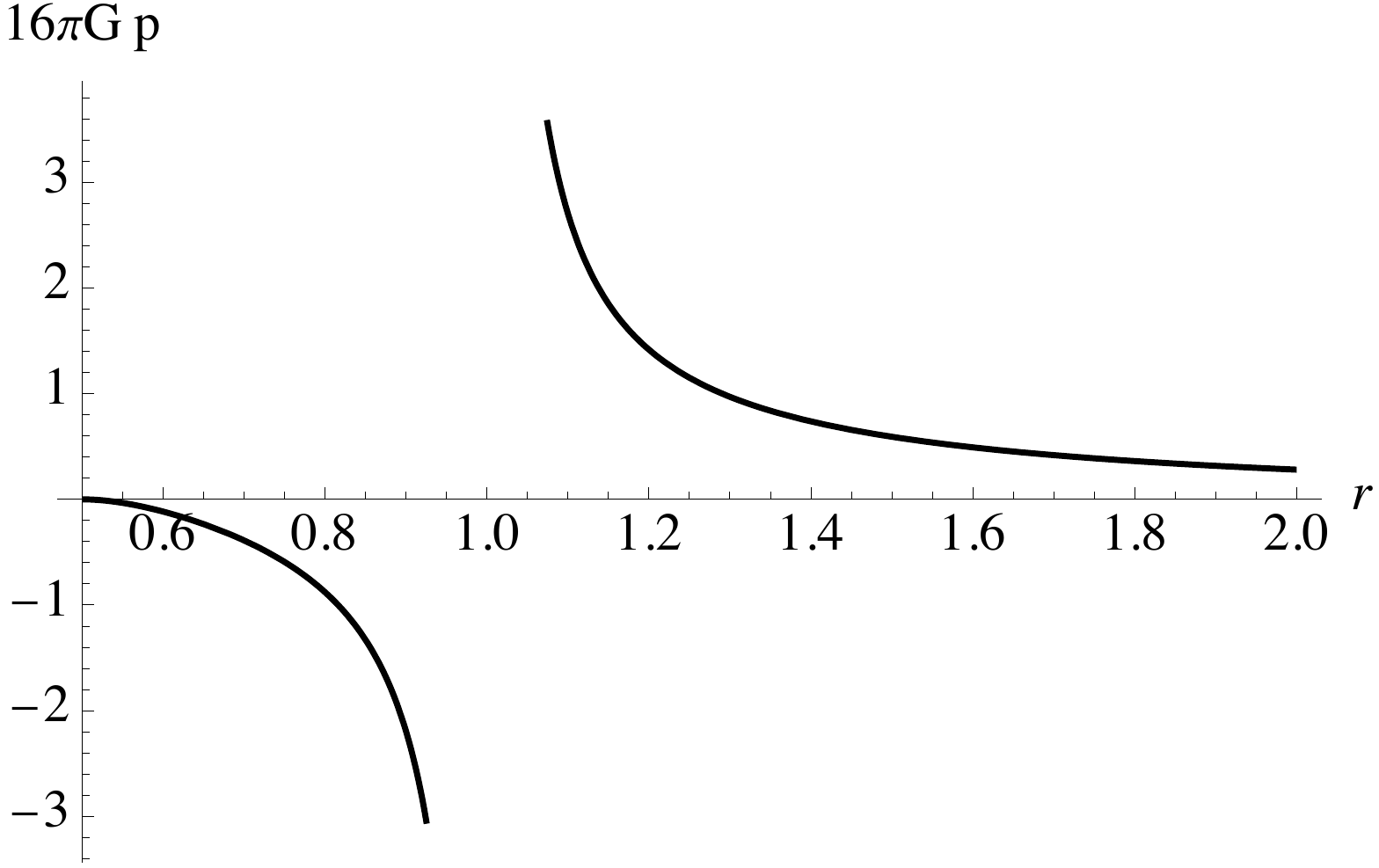}
\caption{The functional dependence of the pressure rescaled by $16\pi G$: from the prograde circular photon orbits in the equatorial plane to the radial infinity (the extremal Kerr BH). The orbits are located on the BH horizon $r^{extr.}_{\cH}=r_s/2=1/2$, where the pressure is equal to zero. The pressure diverges near the double extremal horizon $r_s$ and undergoes the sign flip.}
\label{fig:image1}
\end{figure}

\ssk
Let us take a look at the pressure as the radial coordinate function at $\th=0$ (the north pole). Here, we fix the ``prograde'' and ``retrograde'' photon orbits \rf{photon1}, \rf{photon2} as the reference points. For $r \in [r_1,\infty[$, we get Fig.\ref{fig:image2}, so that $16\pi G p(r,0)$  is a smooth function of the radial direction: from zero on the horizon, having the maximum at $r_s$, and further falling as $1/r^2$. The same qualitative behavior $1/r^2$ is observed for the pressure from the ``retrograde'' photon orbits reference point $r_2=4r^{extr.}_{\cH}$ ($p=60/289 \times 1/16\pi G$ there) to the spatial infinity.

\begin{figure}[ht]
\centering
\captionsetup{width=.8\linewidth}
\includegraphics[width=0.5\linewidth]{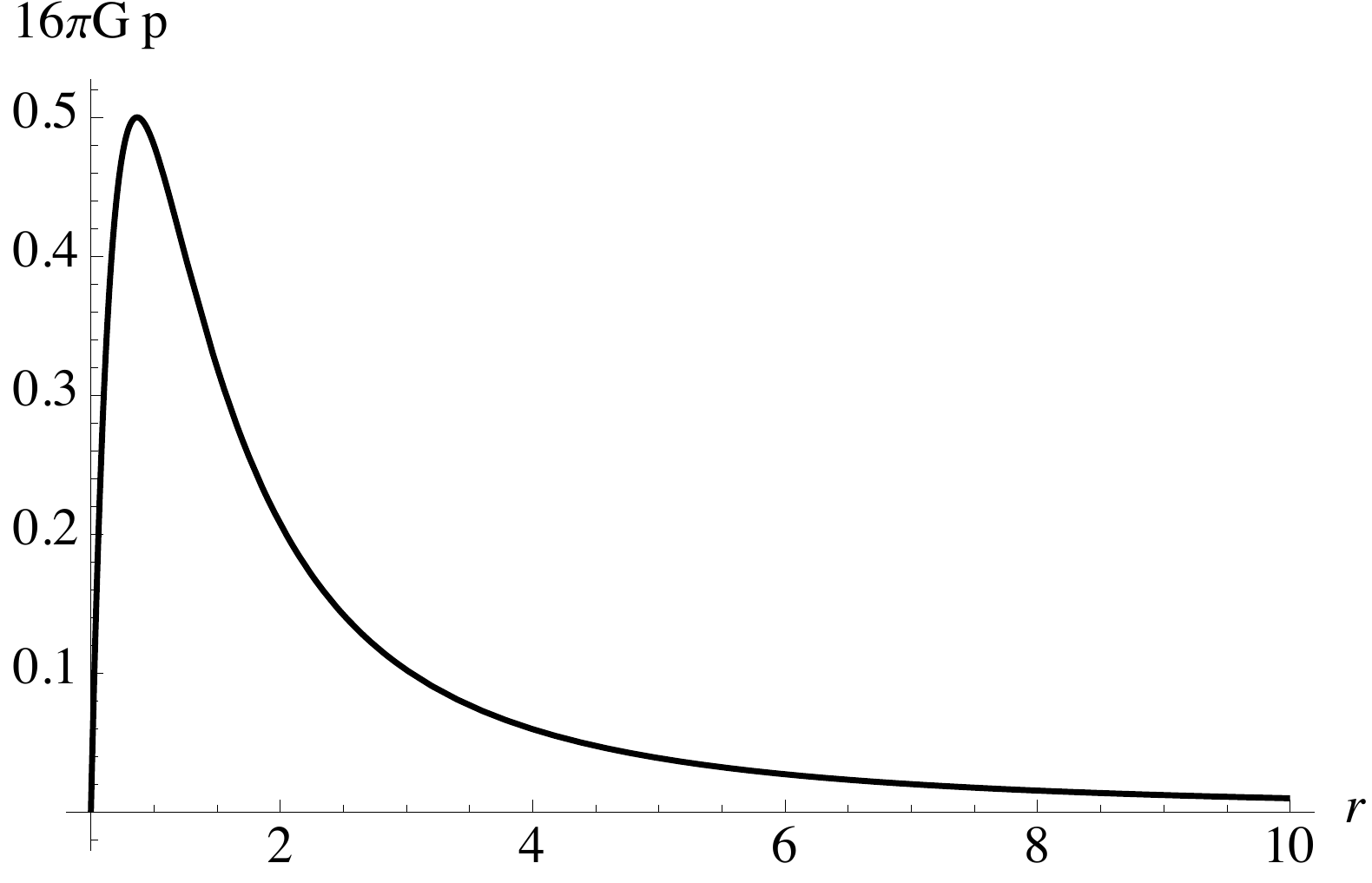}
\caption{The rescaled by $16\pi G$ pressure in the north pole plane (the extremal Kerr BH). The reference point is the BH horizon $r^{extr.}_{\cH}=r_s/2=1/2$, where the pressure is equal to zero.}
\label{fig:image2}
\end{figure}

\ssk
For a slowly rotating black hole, when, say, $C=0.01$, the circular photon orbit radii in the equatorial plane are almost the same. For $r_s=1$, we get $r_1=1.49422$ and $r_2=1.50576$. These values are around $r=3/2 r_s$ which corresponds to the circular photon orbit of the Schwarzschild black hole of the same mass. Plotting the dependence of $16\pi G p$ on the radial coordinate in this case, we observe that for the equatorial/pole planes $16\pi G p$ come to be almost the same: from $16\pi G p \approx 0.45$ at the reference point, falling as $1/r^2$ to the spatial infinity (see Fig.\ref{fig:image3}).

\begin{figure}[ht]
\centering
\captionsetup{width=.8\linewidth}
\includegraphics[width=0.7\linewidth]{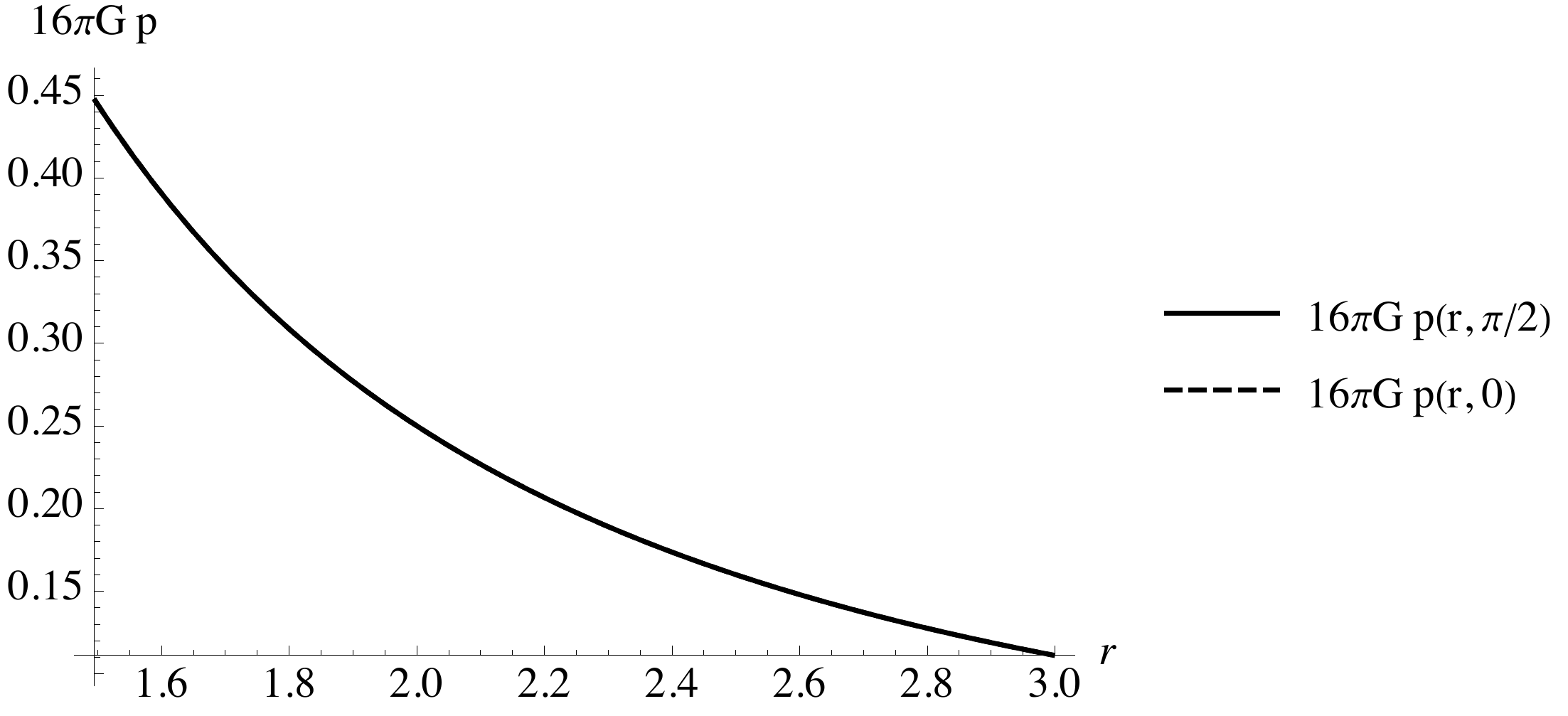}
\caption{The plot of $16\pi G p$ from the prograde circular photon orbits in the equatorial/pole planes of a slowly rotating black hole ($C\equiv a/M=0.01$). In both cases the reference point is  $r_1=1.49422\approx 3r_s/2$.}
\label{fig:image3} 
\end{figure}

\ssk
Now, we turn to the analysis of the expansion \rf{ThKerr2}, which is useful to present as a function of the Schwarzschild radius $r_s=2M$, $C\equiv a/M$ and radial/angle coordinates. It becomes
\[
\Th=\fr{r\Delta}{\r^4(r,\th)}+\fr{C^2 r^4_s r \sin^2\th}{4\r^4(r,\th)\left(\r^2(r,\th)-r_s r \right)}\left(1-\fr{r_s r}{2\r^2(r,\th)}\right)\left(1-\fr{2 r^2}{\r^2(r,\th)} \right)
\]
\be
+\fr1{\r^2(r,\th)}\left(1-\fr{r_s r}{\r^2(r,\th)}\right)\left(r+\fr{C^2 r_s^3 \sin^2\th}{8\r^2(r,\th)} \left(1-\fr{2 r^2}{\r^2(r,\th)} \right) \right),
\la{ThKerr3}
\ee
where
\be
\Delta=r^2-r_s r+\fr14 C^2 r^2_s,\qquad \r^2(r,\th)=r^2+\fr14 C^2 r^2_s \cos^2\th .
\la{DeltarhoCrs}
\ee
The expansion in the equatorial plane of the extremal Kerr BH diverges at the Schwarzschild radius $r_s$ (cf. Fig.\ref{fig:image4}) and undergoes the sign flip at this point. Its behavior is even more interesting at $r>r_s$ since the maximum of the expansion in this domain is reached at $r=2r_s$; after this point the expansion falls as $1/r$ (cf. Fig.\ref{fig:image5}). On the north pole, the expansion becomes a smooth function, reaching the same maximum $\Th=1/2$ as in the previously considered case, and in the same point, $r=2r_s$. It also falls to the spatial infinity as $1/r$ (see Fig.\ref{fig:image6}).

\begin{figure}[ht]
\centering
\captionsetup{width=.8\linewidth}
\includegraphics[width=0.5\linewidth]{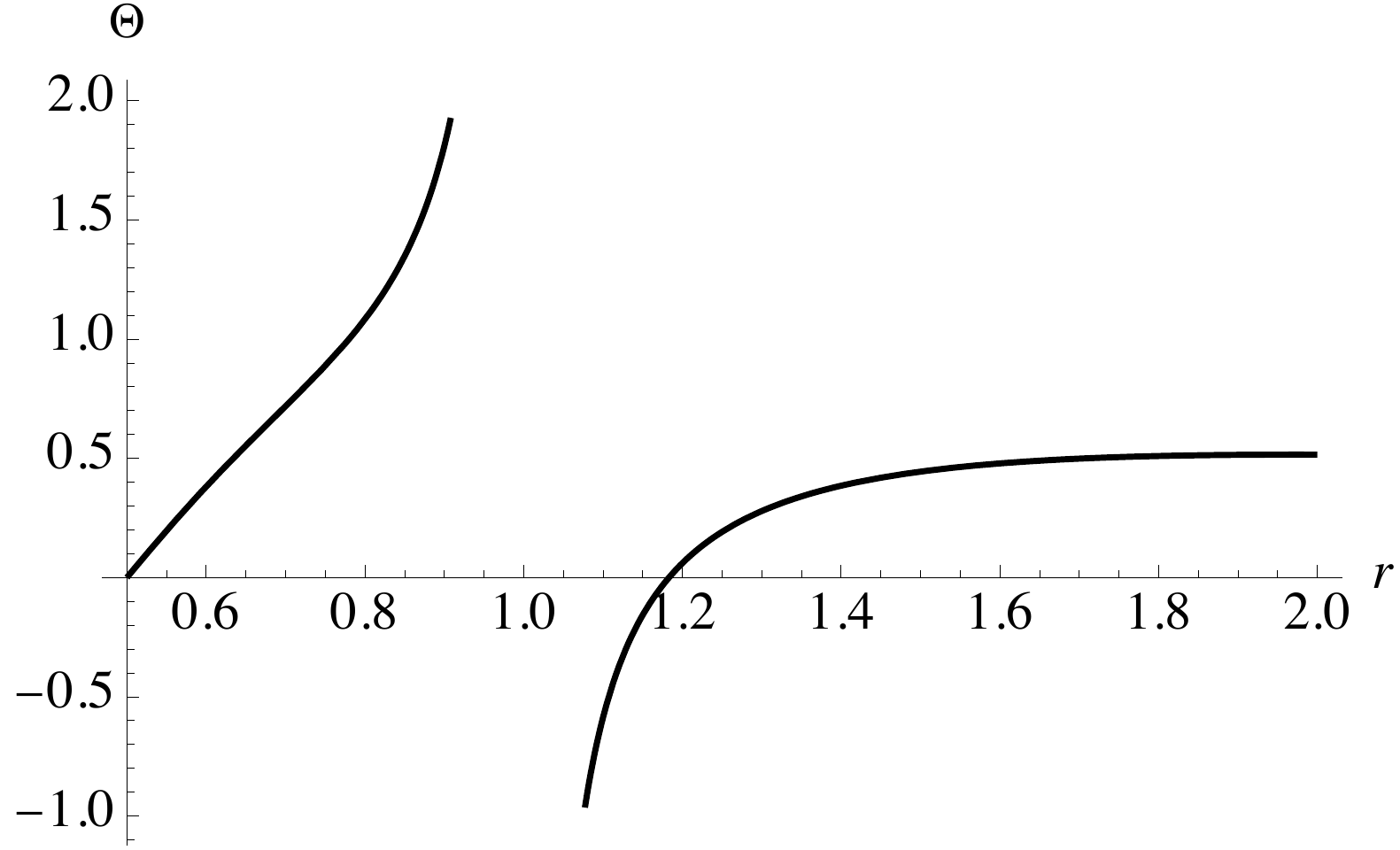}
\caption{The expansion $\Th$ in the equatorial plane of the extremal Kerr BH.}
\label{fig:image4}
\end{figure}

\begin{figure}[ht]
\centering
\captionsetup{width=.8\linewidth}
\includegraphics[width=0.5\linewidth]{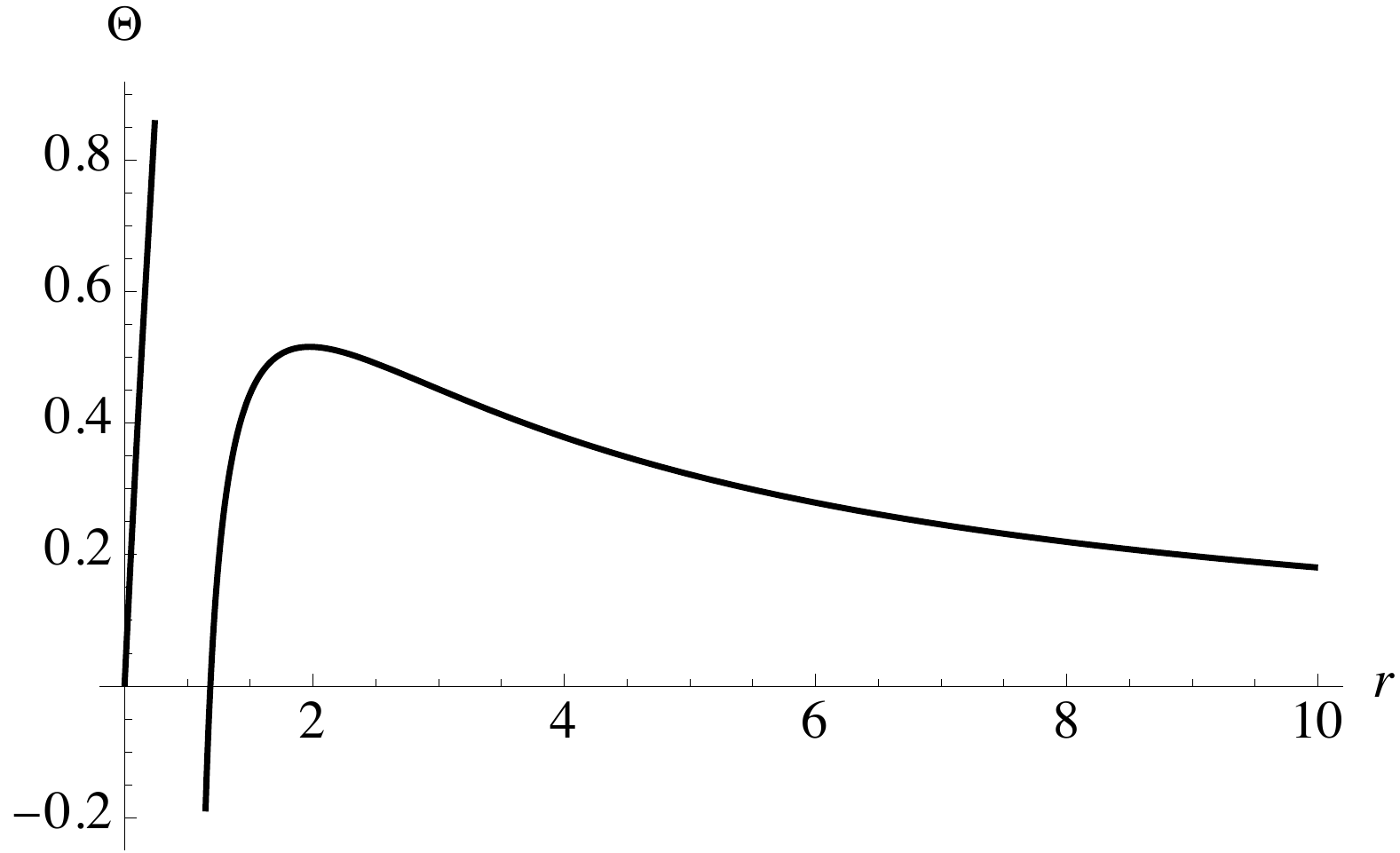}
\caption{The expansion in the equatorial plane of the extremal Kerr BH in an extended domain of the radial coordinate.}
\label{fig:image5}
\end{figure}

\begin{figure}[ht]
\centering
\captionsetup{width=.8\linewidth}
\includegraphics[width=0.5\linewidth]{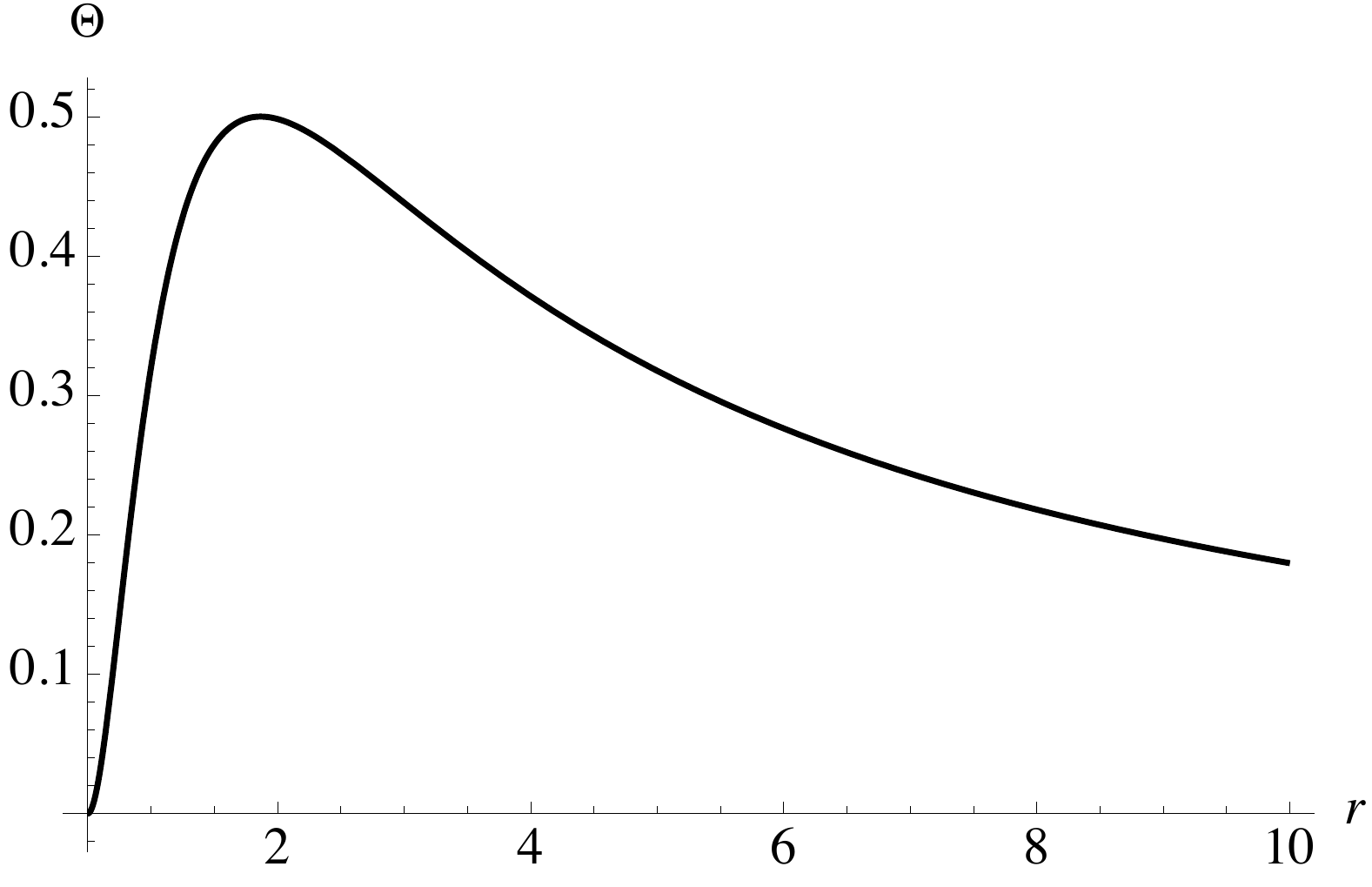}
\caption{The expansion in the north pole plane of the extremal Kerr BH. The reference point is the BH horizon.}
\label{fig:image6}
\end{figure}

\ssk
In the case of the non-extremal Kerr BH with $C=0.01$, we get almost the same functional dependence of the expansion in both -- equatorial and north pole -- planes, when the maximum of $\Th$ is at $r=2r_s$, and it falls to the spatial infinity as $1/r$ (see Fig.\ref{fig:image7}). 

\begin{figure}[ht]
\centering
\captionsetup{width=.8\linewidth}
\includegraphics[width=0.5\linewidth]{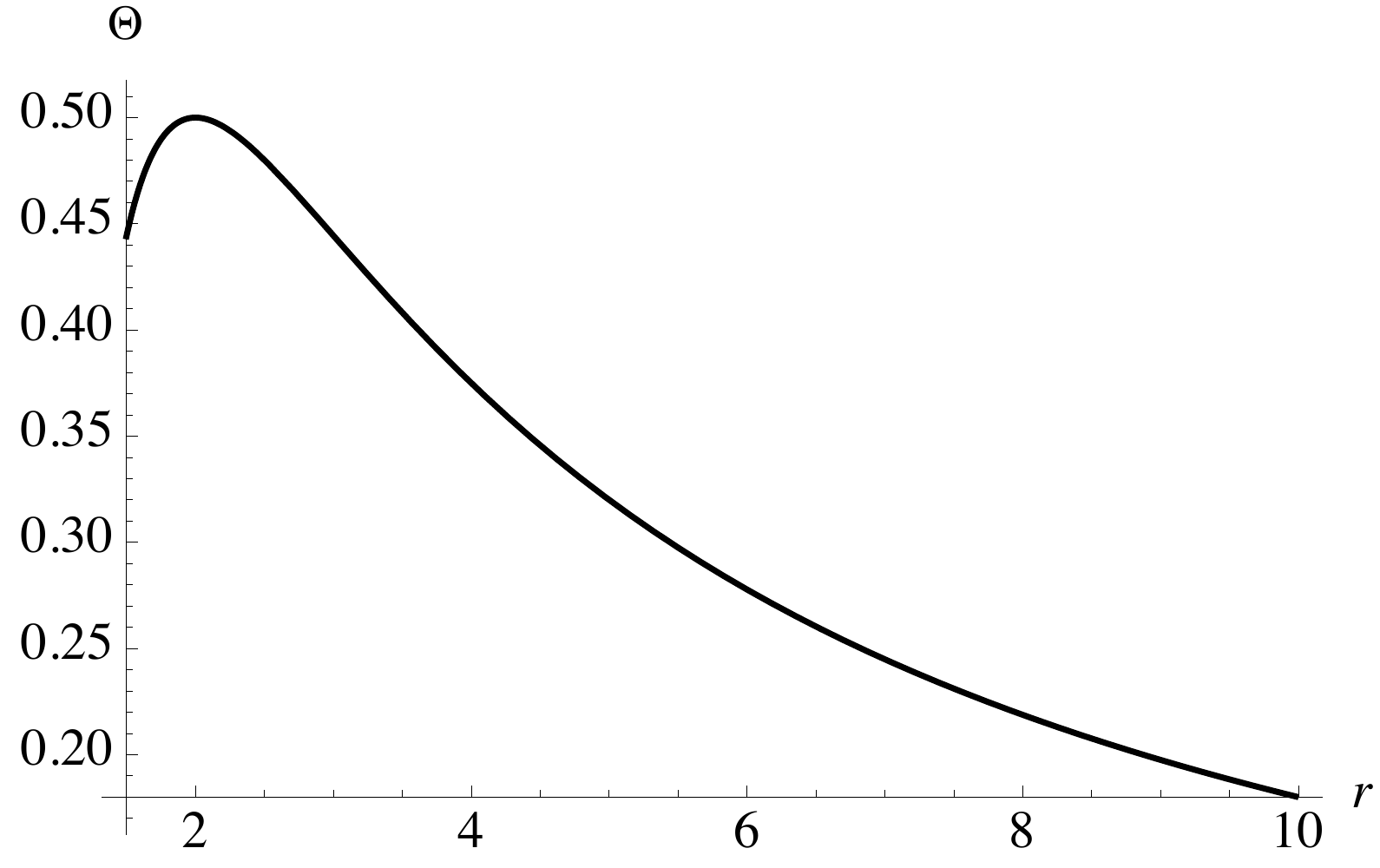}
\caption{The expansion in the equatorial plane of the non-extremal slowly rotating Kerr BH with $C\equiv a/M=0.01$ from the prograde photon orbits location to infinity.}
\label{fig:image7}
\end{figure}

\ssk
For the momentum density $\pi^\vf$, eq. \rf{pivfKerr3}, written in terms of $r_s$ and $C=a/M$,
\be
\pi^\vf(r,\th)=\fr1{8\pi G}\fr{C r^2_s \left(r^2-\fr14 C^2 r^2_s \cos^2 \th \right)}{4\Delta^{1/2}\r^4(r,\th) \left(\r^2(r,\th)-r_s r \right)^{1/2}},
\la{pivfKerr4}
\ee
we obtain the following behavior in both equatorial/pole planes of the extremal Kerr BH, Fig.\ref{fig:image8}. The momentum density diverges at the BH horizon in the pole plane and at the double BH horizon in the equatorial plane. In both cases, $\pi^\vf$ decays as $1/r^4$. For a slowly rotating BH, $\pi^\vf$ is a regular function, falling in both planes to the spatial infinity also as $1/r^4$ (see Fig.\ref{fig:image9}).

\begin{figure}[ht]
\centering
\captionsetup{width=.8\linewidth}
\includegraphics[width=0.7\linewidth]{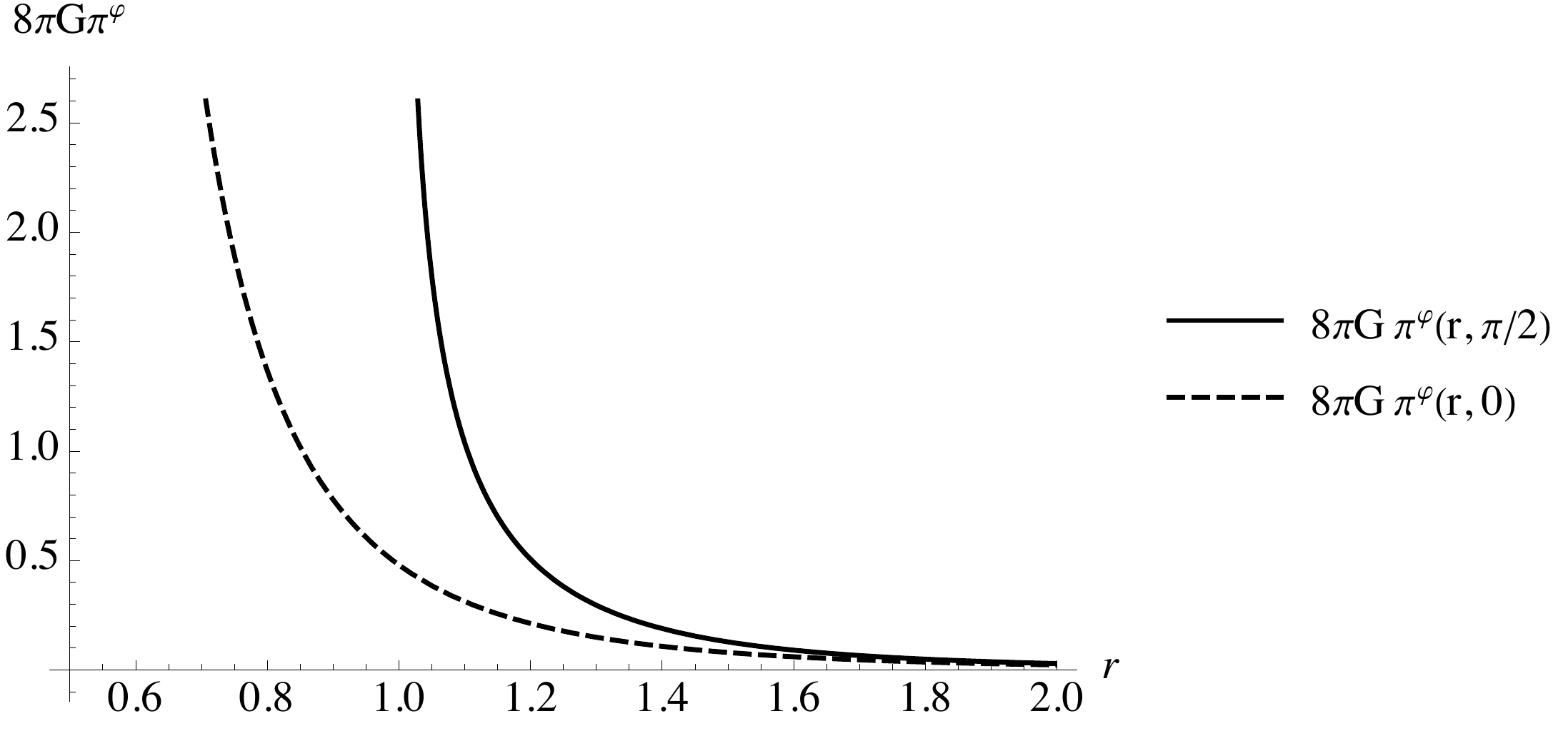}
\caption{The momentum density in the equatorial/pole planes of the extremal Kerr BH.}
\label{fig:image8}
\end{figure}

\begin{figure}[ht]
\centering
\captionsetup{width=.8\linewidth}
\includegraphics[width=0.55\linewidth]{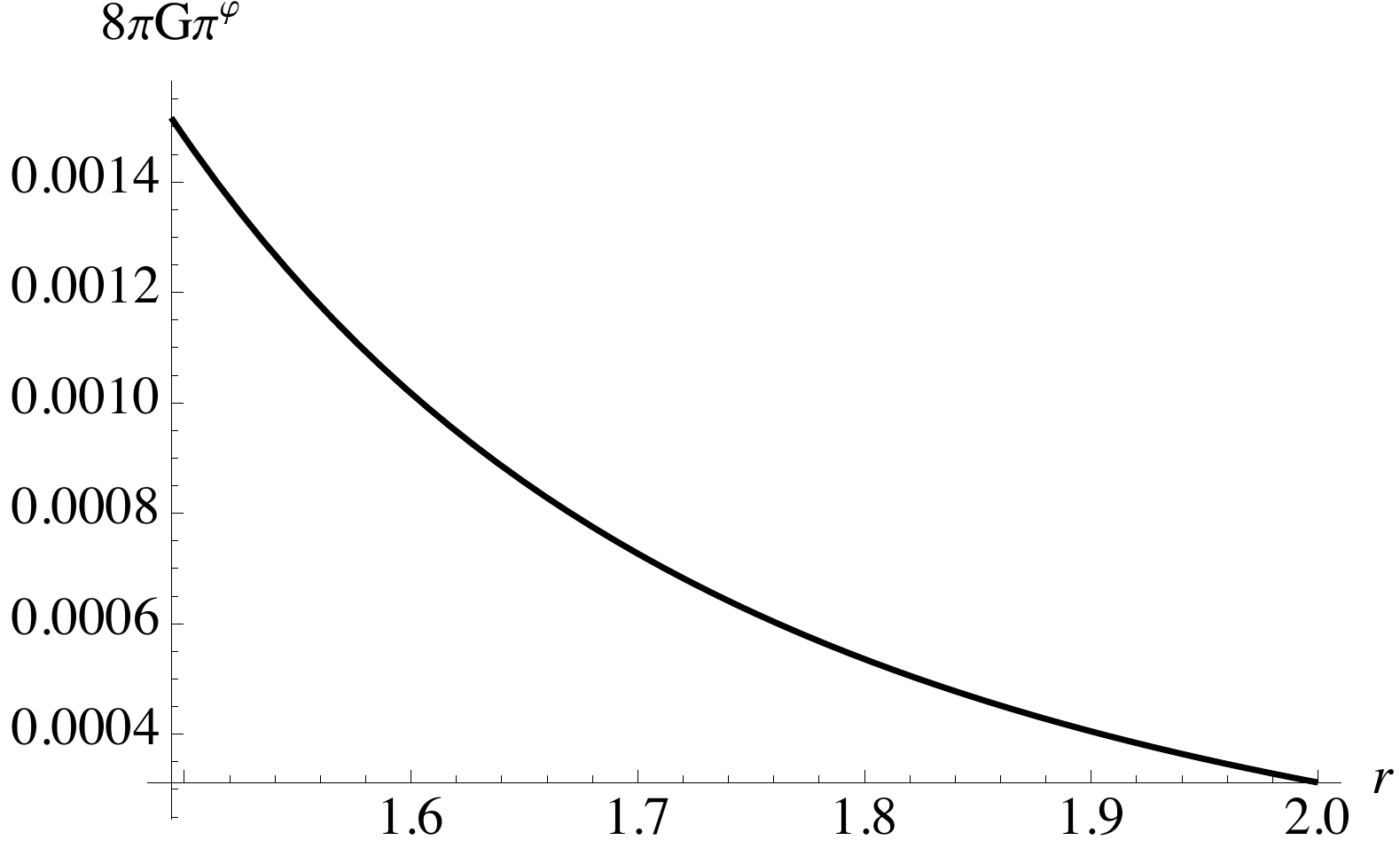}
\caption{The momentum density in the equatorial/pole planes of a non-extremal Kerr BH. ($C\equiv a/M=0.01$; $r$ starts from the ``prograde'' photon orbits.)}
\label{fig:image9}
\end{figure}

\ssk
Finally, the energy density \rf{rhoKerr2} has the following equivalent representation:
\be
\r=\fr1{8\pi G \r^4(r,\th)}\left[\fr{r_s \Delta \left(r^2-\fr14 C^2 r^2_s \cos^2\th \right)}{2 \left(\r^2(r,\th)-r_s r \right)}-\left(\r^2(r,\th)\left(r-\fr12 r_s \right)+r \Delta \right) \right].
\la{rhoKerr3}
\ee 
In the case of the extremal Kerr BH, the functional dependence of the energy density in the equatorial/pole planes is plotted in Fig.\ref{fig:image10}, with the singular point at the Schwarzschild radius in the equatorial plane. For the considered here non-extremal BH  (with $C=0.01$), we get almost coincided regular curves in both planes; see Fig.\ref{fig:image11}.  

\begin{figure}[ht]
\centering
\captionsetup{width=.8\linewidth}
\includegraphics[width=0.65\linewidth]{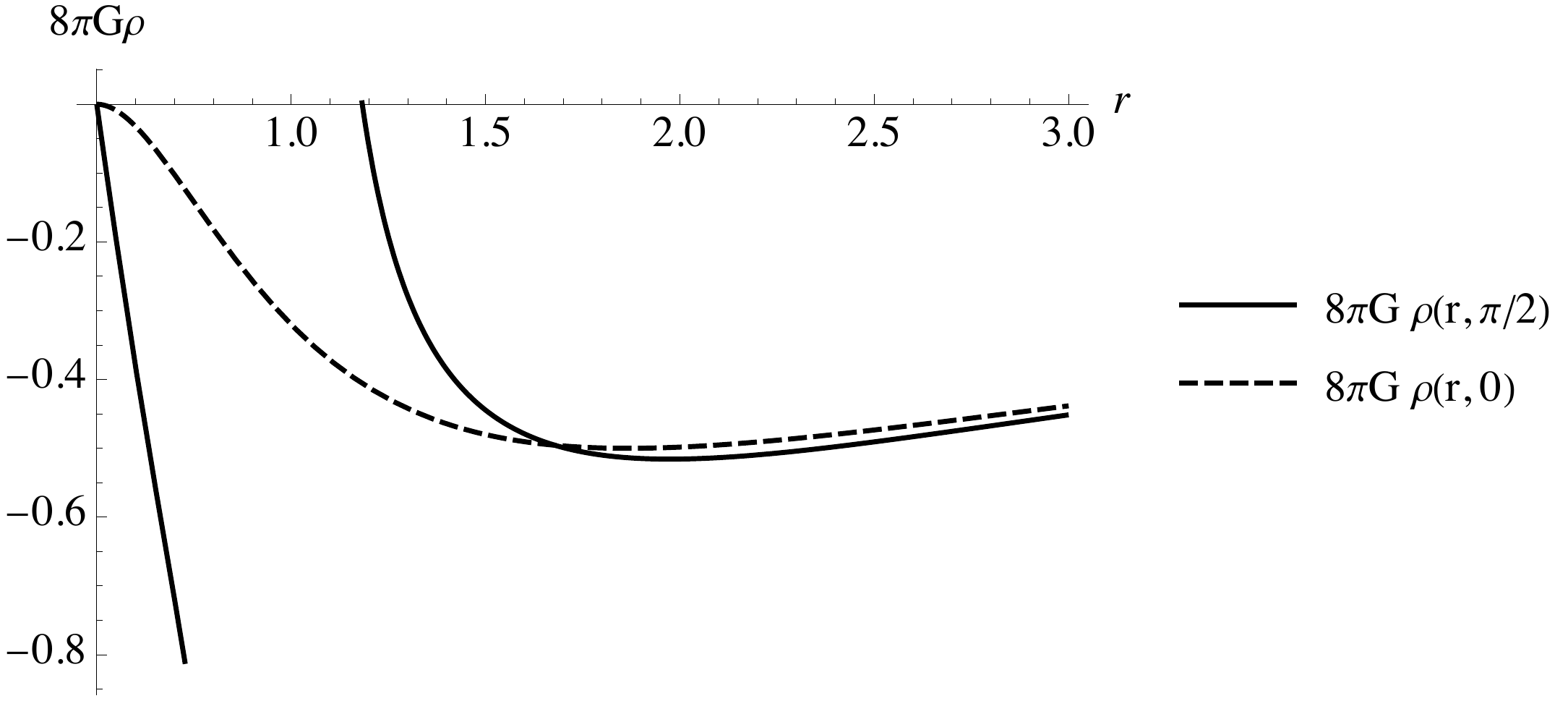}
\caption{The energy density in the equatorial/pole planes of the extremal Kerr BH.}
\label{fig:image10}
\end{figure}

\begin{figure}[ht]
\centering
\captionsetup{width=.8\linewidth}
\includegraphics[width=0.65\linewidth]{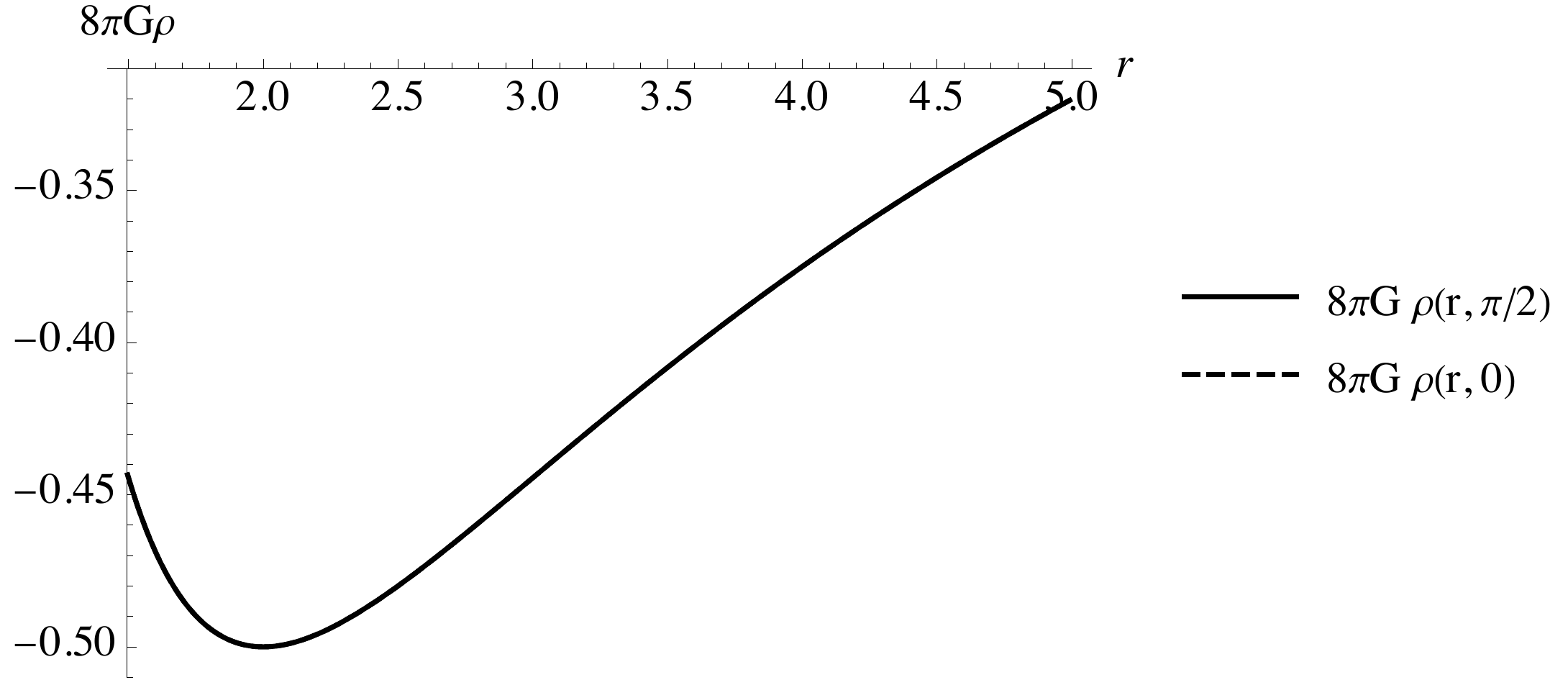}
\caption{The energy density in the equatorial/pole planes of the non-extremal Kerr BH with $C\equiv a/M=0.01$. $r$ starts from the ``prograde'' photon orbits.}
\label{fig:image11}
\end{figure}

\ssk
To summarize, the transport coefficients of the dual fluid $p$, $\Th$, $\r$, and $\pi^\vf$ are complex functions of radial and angle variables. In accordance with the structure of all these functions (cf. eqs. \rf{pKerr4}, \rf{ThKerr3}, \rf{pivfKerr4}, and \rf{rhoKerr3}), they possess the poles defined by $r^2+C^2 r^2_s \cos^2 \th/4-r_s r=0$. These poles may be hidden in specific requirements, an example of which is the transfer of a reference point beyond a singular point. Then, one may find a set of parameters of the Kerr solution (defining the relevant reference point) at which the mentioned coefficients are regular. Generally, we have to take into account the irregularity of these functions.

\section{Summary, discussion, and final remarks}

In this paper, we have applied the Membrane Paradigm to neutral rotating black holes. Following the Parikh-Wilczek approach, well-developed for non-rotating black holes \cite{Parikh:1997ma,Chatterjee:2010gp}, we have constructed the stress-tensor of a stretched membrane, located near the black hole horizon, and to compare it to the stress-tensor of a relativistic viscous fluid. In the nomenclature of the Gauge/Gravity Duality, this effective fluid is dual to the stretched membrane. So Physics on (or in the vicinity of) the black hole horizon is equivalently determined by the corresponding hydrodynamic transport coefficients and susceptibilities of the dual fluid.  

\ssk
In the course of our studies, we have made the following major findings.

\ssk
First, all but two main characteristics of the dual viscous fluid, expressed in terms of geometric variables of the Kerr spacetime, become functions of two variables: the radial coordinate and the polar angle. We have obtained the qualitative behavior of the pressure, expansion, and energy/momentum densities for two specific black holes: the slowly rotating Kerr BH, whose angular momentum is one percent of the black hole mass squared, and the extremal (fast rotating) Kerr BH. Generally, these transport coefficients have poles at different values of the radial coordinate -- in the range between the horizon and the Schwarzschild radius of the black hole -- in dependence on the fixed angle direction. However, if the point of observation is located far from the Schwarzschild radius (this, in particular, happens in the considered here case of the slowly rotating BH with $J=M^2/100$, when the reference point has been chosen to be the photon orbits in the equatorial plane), these functions become smooth in all polar angle planes. It would be interesting to trace back a correspondence of regularity/non-regularity of the transport coefficients on the dual side of the Kerr black hole to the stability of this object, recently proved in \cite{Giorgi:2022omp} and further developed in \cite{Andersson:2022wme}.

\ssk
And second, the transport coefficients -- the shear and bulk viscosities -- remain constants, and we have recovered their values as before (see, e.g., \cite{Damour:1982,Thorne:1986iy,Price:1986yy}). That, in particular, means the famous $\eta/s$ ratio, computed for the most black holes and black branes of String Theory in AdS spacetimes to be $\eta/s \le \hbar/4\pi k_B$ \cite{Kovtun:2003wp}, holds in the case; no violation of the KSS bound \cite{Kovtun:2004de} occurs. (Recall that the issue of violation of this bound was under the focus since establishing this result (see, e.g., \cite{Brigante:2007nu,Erdmenger:2010xm,Jacobson:2011dz,Rebhan:2011vd,Moskalets:2014hoa} for details).) 
Nevertheless, the presence of a negative bulk viscosity still becomes a problem: as it has been noted even in the early works on the Membrane Paradigm (cf. \cite{Thorne:1986iy}), it requires non-causal teleological boundary conditions (see \cite{Bhattacharya:2017dgr} for a review) which means that the horizon fluid is anti-causal in a response \cite{Evans:1996pre}. It does not concern isolated and dynamical horizons \cite{Ashtekar:2004cn,Gourgoulhon:2006uc}, where the response is causal; hence, the bulk viscosity is positive. However, dynamical horizons exist rather for time-dependent solutions than for static/stationary ones. Time-dependent extensions of the Kerr solution developed and studied in \cite{Nurmagambetov:2019mih,Nurmagambetov:2020ann} look perspective to this end.

\ssk
As we have noted at the beginning of the paper, our motivation was to compute main characteristics of the dual viscous fluid within the Membrane Paradigm before starting similar computations on the Gauge/Gravity Duality side. Mainly, our focus was on the computation of the shear viscosity due to importance of this quantity for the Gauge/Gravity Duality. Though we have recovered the classical result of the old Membrane Paradigm, which coincides with the KSS bound coming from the AdS/CFT correspondence, it still looks enigmatic how these different approaches, strictly speaking, unrelated to each other (see \cite{Kovtun:2005ev,deBoer:2014xja} for details), may lead to the same answer. For example, results for 4D bulk viscosity do not look so unambiguous anymore. The AdS/CFT leads to \cite{Gubser:2008sz,Yarom:2009mw}:
\[
\fr{\zeta}{\eta}\ge 2\left(\fr13-c^2_s \right).
\]
Since the physical intuition requires a positive value of the speed of sound square $c^2_s$, possible values of the bulk viscosity are restricted by
\[
\fr{\zeta}{\eta} \le \fr23 \qquad \leadsto \qquad \zeta \le \fr1{24\pi G}\,\,\,\,\,\, \left(\text{for} \,\,\eta=\fr1{16\pi G}\right),
\]
which does not exclude positive values of the bulk viscosity in a holographic fluid.

\ssk
We have reported that the transport coefficients of the extremal Kerr black hole are divergent functions in the radial direction. This result for "internal" degrees of freedom of the BH horizon seems to be in contradiction with the early obtained claims on the regularity of external, w.r.t. the horizon, matter fields for static/stationary solutions of the Einstein equations (see \cite{Nurmagambetov:2019mih,Nurmagambetov:2020ann} and early papers \cite{Park:2017dib,Nurmagambetov:2018het,Nurmagambetov:2019bqz}). However, we have to note that the obtained here results are referred to the Boyer-Lindquist parametrization of the Kerr metric. To avoid truly unphysical singularities, one needs to use the Eddington-Finkelstein parametrization, which is beyond the scope of the paper and the study of which we postpone for future investigations.

\ssk
To sum up, the obtained results, which have shown differences from the case of non-rotating black holes where outcomes of both approaches partially coincide, call for extra theoretical work in figuring out the power of the Membrane Paradigm and its actual correspondence to and differences with the AdS/CFT. Answering to this and other questions will be handy in more deep understanding of microscopic degrees of freedom of a black hole, and what they really are: a condensate of gravitons \cite{Dvali:2012en,Shu:2018isn,Alfaro:2019xye,Dvali:2021ofp}, strings/D-branes \cite{Strominger:1996sh,Brustein:2016msz,Brustein:2021cza}, or other ``molecules''.

\bsk
{\bf Acknowledgements}. We are grateful to all our colleagues around the world for their support during these turbulent times for us. We would also like to express our sincere gratitude to everyone who kindly help Ukrainian refugees forced to leave their homes. AJN personally thanks Leonid and Tetyana Dudnik for their warm hospitality during the leave of Kharkiv.

\newpage

\appendix
\numberwithin{equation}{section}

\section{Details of computations: stress tensors of stretched membrane in Schwarzschild geometry and its dual viscous fluid}

According to the Membrane Paradigm \cite{Thorne:1986iy,Price:1986yy,Suen:1988kq}, the membrane stress tensor in 1+1+2 coordinate split comes as follows
\cite{Parikh:1997ma}:
\be
t_{ab}=\fr1{8\pi G} \left(K h_{ab}-K_{ab} \right),\qquad h_{ab}=\g_{ab}-u_a u_b ,
\la{tabdef}
\ee
where $h_{ab}$ is the metric tensor of a 3D hypersurface orthogonal to the space-like unit vector $n^a$ (radial direction for the Schwarzschild); $\g_{ab}$ is the metric of a 2D hypersurface orthogonal as to $n^a$, as well as to the time-like vector $u^a$. $K_{ab}$ is the extrinsic curvature of the 3D hypersurface; $K$ is the trace of $K_{ab}$.

The stretched membrane stress tensor \rf{tabdef} can be equivalently represented by the stress-tensor of a viscous fluid \cite{RezzollaBook}:
\be
t_{ab}=\fr1{\a} \r u_a u_b+\fr1{\a}\g_{aA}\g_{bB} \left(p \g^{AB}-2\eta \s^{AB}-\zeta \Th \g^{AB} \right)+\pi^A\left(\g_{aA}u_b+\g_{bA}u_a \right).
\la{tabfluiddef}
\ee
Recall that indices $a,b$ are 4-dimensional indices; indices $A,B$ are two-dimensional indices of a surface parameterized by angles $\{\th,\vf\}$. In \rf{tabfluiddef}, $\r$ is the energy density, $p$ is the fluid pressure, $\s_{ab}$, and $\Th$ are the shear tensor and the expansion of the null geodesics near the event horizon. Also, $\eta$ is the shear viscosity, $\zeta$ is the bulk viscosity, and $\pi^A$ is the momentum density. In particular, note the presence of the renormalization parameter $\a$, the role of which is to make the divergent on the true horizon quantities finite (see \cite{Parikh:1997ma} and further discussion for details). 

\ssk
Let us review how eq. \rf{tabdef} can be used to read off various characteristics of an effective fluid dynamics from eq. \rf{tabfluiddef} for the Schwarzschild solution.

\ssk
The Schwarzschild geometry is determined by 
\be
ds^2 = -f(r)dt^2 + \frac{dr^2}{f(r)} + r^2\left(d\theta^2 + \sin^2\theta d\varphi^2\right),\qquad f(r)=1-\fr{2M}{r} .
\la{SS}
\ee
We are aimed at splitting this metric into $1+1+2$ form by the use of time-like and space-like vectors $u^a$ and $n^a$ near the horizon, turned into the same null-vector $l^a$ on the horizon. This split takes the form
\be
ds^2 = -u_a u_b \,dx^adx^b + n_a n_b \,dx^adx^b + \gamma_{ab}\,dx^a dx^b,
\la{SS112}
\ee
where
\be
\begin{split}
&u_a dx^a = f^{1/2}dt\rightarrow u_a = f^{1/2}\d_a^t\\
&n_a dx^a = f^{-1/2}dr \rightarrow n_a = f^{-1/2}\d_a^r\\
\g_{ab}dx^a dx^b = r^2d\theta^2 + r^2\sin^2\theta d\varphi^2&\rightarrow \g_{AB} = \begin{pmatrix}r^2 && 0\\
0&& r^2\sin^2\theta\end{pmatrix}\;\;A, B = \{\theta, \varphi\} .
\end{split}
\la{ungamSS}
\ee
Therefore, in the dual to 1-forms basis,
\be
u^a=-f^{-1/2}\pa_t,\qquad n^a=f^{1/2} \pa_r,
\la{undual}
\ee
so that $u^a u_a=-1$, $n^a n_a=1$, $u^a n_a=0$, and $n^b \nabla_b n_a=0$. The latter relation justifies the form of the stress tensor \rf{tabdef} (see details in \cite{Parikh:1997ma}).

\ssk
To recover \rf{tabdef} in the background \rf{SS}, we have to compute the extrinsic curvature of a stretched horizon (a 3-dimensional surface orthogonal to $n_a$) and its trace. For the extrinsic curvature tensor, we have
\be
K_{ab} = \nabla_a n_b = -f^{1/2}\begin{pmatrix}\frac12\partial_rf&&0&&0&&0\\
0&&0&&0&&0\\
0&&0&&-r&&0\\
0&&0&&0&&-r\sin^2\theta
\end{pmatrix} = -\frac12f^{-1/2}\partial_rfu_au_b + \frac{f^{1/2}}{r}\g_{ab},
\la{KabSS}
\ee
and the trace becomes
\be
K = K^a_a = \frac{1}{2}f^{-1/2}\partial_rf + \frac{2f^{1/2}}{r}.
\la{KSS}
\ee

\ssk
Within the Parikh-Wilczek aproach \cite{Parikh:1997ma}, the extrinsic curvature of the stretched horizon is replaced accordingly with 
\be
K_{ab} \ra \a^{-1} k_{ab}-\a^{-1} g_{\cH} u_a u_b .
\la{Kabstrhor}
\ee
Technically, in \rf{Kabstrhor}, we further split $K_{ab}$ onto a two-dimensional surface (orthogonal to $u_a$ of \rf{ungamSS}) extrinsic curvature $k_{AB}$ (then $k_{ab}=\g_{aA}\g_{bB}k^{AB}$) and the surface gravity $g_{\cH}$.  At this point, one may wonder how the extrinsic curvature of the orthogonal to $u^a$ 2D hypersurface, {\it lying completely inside} the 3D hypersurface, is related to the extrinsic curvature tensor $K_{ab}$, defining the geometry of the hypersurface normal to a vector ($n^a$ in the case) directed {
\it outside}. This match happens due to a coincidence of $u^a$ and $n^a$ vectors in the null limit when $\a u^a \ra l^a$ and $\a n^a \ra l^a$ at $\a \ra 0$. Here, $\a$ is the renormalization factor, needed to keep finiteness of $K_{ab}$ components on the horizon, and $l^a$ is the null vector normal and tangential to the horizon. The renormalization factor $\a$ is chosen as $\a=f^{1/2}$ in the case.\footnote{The requirement for $\a$ comes as follows \cite{Parikh:1997ma,Chatterjee:2010gp}. In the $\a \ra  0$ limit, $\a u^a \ra l^a$ and $\a n^a \ra l^a$, where $l^a$ is a null vector ($l^a l_a=0$) which is tangential and normal to the horizon simultaneously. Clearly, the specific representation of such a null vector depends on the choice. The norm of $\a u^a$ and $\a n^a$ of \rf{undual} with $\a=f^{1/2}$ is equal to $f$ which becomes zero on the horizon.}

\ssk
Indeed, in the limit $\a \ra 0$, when the stretched membrane turns into the true event horizon surface (null hypersurface), the trace of $K_{ab}$ (cf. \rf{KSS}) diverges due to the simple pole from $f(r)$ at $r=r_{\cH}$. In this limit,
\be
\lim_{\a \ra 0} K=\fr12\,f^{-1/2}\pa_r f\Big|_{r=r_{\cH}} \sim \Tr \left(\a^{-1} k_{ab}-\a^{-1} g_{\cH} u_a u_b \right)\Big|_{r=r_{\cH}},
\la{Knull}
\ee
hence choosing $\a \sim f^{1/2}$ makes geometric quantities regular on the horizon. The same concerns components of $K_{ab}$ (see \rf{KabSS}). $K_{tt}=K_{ab}u^a u^b$ clearly diverges in the null limit:
\be
\lim_{\a \ra 0} K_{tt}=-\fr12 f^{-1/2} \pa_r f \Big|_{r=r_{\cH}} \longrightarrow -\a^{-1}g_{\cH} \Big|_{r=r_{\cH}}, 
\la{Kttnull}
\ee
but fixing the renormalization parameter $\a \sim f^{1/2}$ results in the correct expression for the surface gravity on the horizon (see eq. \rf{ThsiggHSS} in below).

\ssk
Next, we can divide the 2-dimensional extrinsic curvature $k_{AB}=k_{BA}$ onto the traceless part and the corresponding trace $\Th=\g^{AB}k_{AB}$:
\be
k_{AB}=\s_{AB}+\fr12 \Th \g_{AB},\qquad \Tr \,\s_{AB}=0 .
\la{kABsplit}
\ee
Hence, inserting \rf{Kabstrhor} and \rf{kABsplit} into the stress tensor \rf{tabdef}, we arrive at
\be
t_{ab}=\fr1{8\pi G} \left(\left(\a^{-1}\Th+\a^{-1}g_{\cH} \right)\left(\g_{ab}-u_a u_b \right)+\a^{-1} g_{\cH} u_a u_b-\a^{-1} \left(\s_{ab}+\fr12 \Th \g_{ab} \right)\right).
\la{tabKsplit}
\ee
And after some algebra we get
\be
t_{ab}=\fr1{8\pi G \a} \left[-\Th\, u_a u_b-\s_{ab}+\left(\fr12 \Th+g_{\cH} \right) \g_{ab} \right].
\la{tabKsplitfin}
\ee

\ssk
Now, we can compare stress tensors \rf{tabKsplitfin} and \rf{tabfluiddef} with the following outcome:
\[
\r=-\fr1{8\pi G} \Th,\qquad 2\eta=\fr1{8\pi G},
\]
\[
p-\zeta \Th=\fr1{8\pi G} \left(\fr12 \Th+g_{\cH} \right).
\]
So we get the following correspondence between characteristics of the dual fluid and properties of the stretched membrane in the specific geometric background:
\be
\r =-\fr1{8\pi G} \Th, \qquad \eta=\fr1{16\pi G},
\la{rhoeta}
\ee
\be
p=\fr{g_{\cH}}{8\pi G},\qquad \zeta=-\fr1{16\pi G} .
\la{pzeta}
\ee
Also, the momentum density in angular directions is $\pi^A=0$.

\ssk
The other task is to relate $\Th$, $g_{\cH}$, and $\s_{AB}$ to the geometric configuration of the space-time in hands. We have to compare the stress tensor \rf{tabKsplitfin} to $t_{ab}=(K h_{ab}-K_{ab})/8\pi G$, which is (cf. \rf{KabSS}, \rf{KSS})
\be
t_{ab} =\frac{1}{8\pi G}\left[\left(\frac{1}{2f^{1/2}}\partial_rf + \frac{f^{1/2}}{r}\right)\g_{ab} - \frac{2f^{1/2}}{r}u_au_b\right].
\la{tabKKabSS}
\ee
We obtain
\be
\Th=\fr{2f(r)}{r},\qquad \s_{AB}=0,\qquad g_{\cH}=\fr{\pa_r f(r)}{2}.
\la{ThsiggHSS}
\ee
Hence, the shear tensor and the expansion are trivial on the black hole horizon, and the surface gravity is finite there and is equal to 
$g_{\cH}=(4M)^{-1}$ as it should be.\footnote{The fact of finiteness of the stress-tesor components near the horizon of static and stationary black holes even on account of quantum corrections to gravity action was pointed out in \cite{Park:2017dib,Nurmagambetov:2018het,Nurmagambetov:2019bqz,Nurmagambetov:2019mih,Nurmagambetov:2020ann}.} 

\ssk
Additionally, one may check the relevance of \rf{ThsiggHSS} computing the 2-dimensional extrinsic curvature $k_{AB}$. By definition, 
\be
k_{AB}=\fr12 \cL_l \g_{AB},
\la{kABdef}
\ee
where $l^a$ is the null-vector on the horizon, determined by $l^a=\a n^a$ (recall, $\a=f^{1/2}$), and $\cL_l$ is the Lie derivative along the null-vector. Then, the $2\times 2$ angular part of the full $K_{ab}$ tensor becomes
\be
k_{AB}=\left(
\begin{array}{cc}
r f &0\\
0& rf \sin^2\th
\end{array}
\right)=\fr12 \left(\fr{2f}{r} \right) \g_{AB}.
\la{kABcomp}
\ee
Comparing this result to the general structure of $k_{AB}$, eq. \rf{kABsplit}, we arrive at
\be
\s_{AB}=0,\qquad \Th=\fr{2f}{r} 
\la{sThkAB}
\ee
again.

\ssk
For the Schwarzschild geometry, we get the trivial (everywhere in the acceptable domain of coordinates) value of the shear tensor. One may wonder how we have got the value of the shear viscosity \rf{rhoeta} if $\s_{ab}$ is presented in \rf{tabKsplitfin} just formally. The value of the shear viscosity can not be fixed from $\eta \s_{ab}=0$ equation in this case, which requires taking additional arguments. For instance, it could be (projected) Einstein equations which can be written in the form of Navier-Stokes-like equations (Damour-Navier-Stokes (DNS) equations) \cite{Damour:1982,Damour:2008ji,Padmanabhan:2010rp}. Then, the effective shear viscosity can be extracted from the structure of the DNS equations. 

\ssk
Now, we briefly describe the reason for using the standard stress-tensor \rf{tabdef} (or eq. \rf{STtensor} in the main text) in the case of non-trivial acceleration of $n_a$ defined by $a^c=n^b \nabla_b n^c$. Here, we follow \cite{Li:2017ljz} and slightly correct the result mentioned therein. The relevant part of the gravitational action variation that includes the $n_a$ field acceleration is (cf. eq. (A1) of \cite{Parikh:1997ma}):
\be
\d S_{\text{out}}\Big|_{a^c \ne 0}=\int d^3x\,\sqrt{-h}\left[\left(a_a n^a h^{bc}+n^c a^b+n^b a^c\right)\d h_{bc}- h^{bc} a_c n^a \d h_{ab}-n^a a^b \d h_{ab}\right].
\la{dSant}
\ee 
The difference in \rf{dSant} and eq. (A1) of \cite{Li:2017ljz} is in the term $a_a n^a h^{bc} \d h_{bc}$ which is equal to zero, due to $n^a a_a=0$. Then, following directly \cite{Li:2017ljz}, we can use the symmetry of $\d h_{ab}$ and further write \rf{dSant} as
\[
\d S_{\text{out}}\Big|_{a^c \ne 0}=\int d^3x\,\sqrt{-h}\left(n^a a^b -h^{bc} a_c n^a\right)\d h_{ab}= \int d^3x\,\sqrt{-h}\left(g^{bc}-h^{bc} \right)n^a a_c \d h_{ab}
\]
\be
=\int d^3x\,\sqrt{-h}\, n^a n^b n^c a_c \d h_{ab}.
\la{dSant1}
\ee
The variation is equal to zero due to $n^c a_c=0$ again.

\section{Dimension analysis}

To control the correctness of different quantities, it is useful to make the dimension analysis of variables. Since the metric tensor is chosen to be dimensionless, different quantities entering line elements have different dimensions in length. The line element by itself has the dimension $l^2$. Then, $[dt^2]=l^2=[dr^2]$, spherical angles are dimensionless -- $[d\th^2]=l^0=[d\vf^2]$ -- so that the entities in the Kerr metric \rf{Kerrds2} get received
\be
[F_t]=l^0,\qquad [F_r]=l^0,\qquad [f]=l,\qquad [\r^2(r,\th)]=l^2, \qquad [F_\vf]=l.
\la{Kerrds2dims}
\ee
As a result, the BH mass has the dimension of length: $[M]=l$. And different components of the 3D extrinsic curvature tensor $K_{ab}$ have different dimensions:
\be
[K_{tt}]=l^{-1},\qquad [K_{t\vf}]=l^0,\qquad [K_{\th\th}]=l,\qquad [K_{\vf\vf}]=l.
\la{KabdimsKerr}
\ee
The latter in particulr means that
\be
[\s_{\th\th}]=[\s_{\vf\vf}]=l,\qquad [\Th]=l^{-1},\qquad [\g_{AB}]=l^2.
\la{ThdimKerr}
\ee
From the exact form of $\r^2(r,\th)$, it is clear that $[a]=l$; hence, the dimension of $J=Ma$ is $[J]=l^2$.


\end{document}